\documentclass[acmsmall,screen=true,nonacm]{acmart}

\AtBeginDocument{%
  \providecommand\BibTeX{{%
    \normalfont B\kern-0.5em{\scshape i\kern-0.25em b}\kern-0.8em\TeX}}}

\setcopyright{acmcopyright}
\copyrightyear{2018}
\acmYear{2018}
\acmDOI{10.1145/1122445.1122456}

\acmConference[Woodstock '18]{Woodstock '18: ACM Symposium on Neural
  Gaze Detection}{June 03--05, 2018}{Woodstock, NY}
\acmBooktitle{Woodstock '18: ACM Symposium on Neural Gaze Detection,
  June 03--05, 2018, Woodstock, NY}
\acmPrice{15.00}
\acmISBN{978-1-4503-XXXX-X/18/06}

\usepackage{CJK}
\usepackage[caption=true]{subfig}
\usepackage{verbatim}
\usepackage[title]{appendix}
\usepackage{multirow}

\usepackage{graphicx}
\usepackage[normalem]{ulem}
\useunder{\uline}{\ul}{}
\usepackage{lscape}

\begin{document}
\begin{CJK}{UTF8}{gbsn}

\title[Malicious Selling Strategies in E-Commerce Livestream]{Malicious Selling Strategies in E-Commerce Livestream: A Case Study of Alibaba's Taobao and ByteDance's TikTok}
\author{Qunfang Wu}
\affiliation{%
 \institution{Syracuse University}
 \city{Syracuse}
 \state{New York}
 \country{USA}}

\author{Yisi Sang}
\affiliation{%
 \institution{Syracuse University}
 \city{Syracuse}
 \state{New York}
 \country{USA}}

\author{Dakuo Wang}
\affiliation{%
  \institution{IBM Research}
  \city{Boston}
  \state{Massachusetts}
  \country{USA}}

\author{Zhicong Lu}
\authornote{Corresponding author.}
\affiliation{%
  \institution{City University of Hong Kong}
  \city{Hong Kong}
  \country{China}}
\email{calebluzc08@gmail.com}
\renewcommand{\shortauthors}{Wu, et al.}


\begin{abstract}
Due to the limitations imposed by the COVID-19 pandemic, many users have shifted their shopping patterns from offline to online. Livestream shopping has become popular as one of the online shopping media. However, many streamers' malicious selling behaviors have been reported. In this research, we sought to explore streamers' malicious selling strategies and understand how viewers perceive these strategies. First, we recorded 40 livestream shopping sessions from two popular livestream platforms in China---Taobao and TikTok (or ``Douyin'' in Chinese). We identified 16 malicious selling strategies and found that platform designs enhanced these malicious selling strategies. Second, through an interview study with 13 viewers, we provide a rich description of viewers' awareness of malicious selling strategies and the challenges they encountered while trying to overcome malicious selling. We conclude by discussing the policy and design implications of countering malicious selling.
\end{abstract}



\maketitle

\section{Introduction}
Livestream shopping has become a popular, new form of online shopping~\cite{walmart2020tiktok,cai2019live}.
In a typical 30-minute to one-hour livestreaming session, the streamer, who is often a shop owner, model, influencer, or celebrity, spends a few minutes introducing each product (e.g., food or clothes) and then encourages their viewers to place an order for it right away. Livestreaming platforms have an integrated ``shopping window'' feature, whereby whenever the streamer goes over the product one by one, viewers can see a purchase link for that product pop up on their screen.
They can then make a purchase within the livestream with a simple click without needing to enter address or credit card information. 
Despite seeming similar to Quality Value Convenience or the Home Shopping Network\footnote{``Quality Value Convenience'' and ``Home Shopping Network'' are American free-to-air televised home shopping channels.} on television, livestream shopping offers viewers a more interactive and engaging experience~\cite{benedicktus2010conveying,kpmg2020}.
Hundreds of thousands of viewers can watch a livestream session simultaneously and directly interact with the streamer via a built-in chat channel or tip the streamer streamer with a virtual gift (e.g., a flower).
Some viewers may eventually make a purchase, while others will just enjoy watching the streamer chatting with their viewers. 

Since 2019, livestreaming-based e-commerce has taken off in China~\cite{kapner2020qvc}. This new form of online shopping benefits from ``the increasing tensions [that exist] between the cultural politics and economic ambitions of digital China''~\cite{cunningham2019china}.
The government encourages small businesses to sell their products via livestreams, which enables small businesses to reach more customers at a lower cost. Moreover, the COVID-19 pandemic fueled the livestream shopping frenzy, as home-bound customers could not go to brick and mortar stores to shop so retailers had to rely on online shopping platforms for sales.
Platforms that support livestream and e-commerce (e.g., Alibaba's\footnote{Alibaba is a Chinese Technology company, specializing in e-commerce, retail services, cloud computing, online payments, etc.} Taobao, ByteDance's TikTok, or Douyin in Chinese) have integrated a range of features to promote engagement and purchases, e.g., red pockets\footnote{A ``red pocket,'' or known as ``red envelope,'' is a monetary gift given during holidays or weddings in East and Southeast Asian cultures. It is named after its outer packing, usually taking the form of a red envelope or red pocket. The sending of digital red pockets became popular in the 2010's due to the emergence of mobile wallet systems in messaging apps.}, fan groups, and loyalty levels~\cite{liu2021learn}. These integrated features enable e-commerce applications to support more social activities and facilitate communication and connections between businesses and customers. 
Compared to China, livestream commerce in the U.S. and other countries is nascent in terms of platform designs and services. E-commerce platforms (e.g., Amazon) and social media platforms (e.g., Facebook Live) are catching up with trend~\cite{wongkitrungrueng2020live,walmart2020tiktok}.

This rapid development of livestream e-commerce in China, however, does comes with a cost. More and more malicious selling incidents have been reported by the media, such as selling fake products~\cite{insights2020xinba} and tax evasion~\cite{cnn2021viya}. Recent studies have investigated the motivations of livestream viewers' purchasing behavior and of streamers' selling behavior~\cite{cai2018utilitarian,xu2020drives,li2020understanding,cai2019live,su2019empirical}, but the negative side of livestream shopping, i.e., the malicious selling strategies used by streamers, and the platform designs that mediate these strategies, is underexplored. 

This research defined \textit{livestream malicious selling} as streamers leveraging unbalanced power and information to deceive, coerce, or manipulate viewers into buying behavior that is against viewers' best interests during livestream sessions. To understand malicious selling on livestream shopping platforms, we ask three research quesstions. \textbf{RQ1:} What malicious selling strategies are used by streamers on livestream shopping platforms? \textbf{RQ2:} How do platform designs mediate these malicious selling strategies? \textbf{RQ3:} How do viewers perceive (i.e., awareness and attitude) and react to these malicious selling strategies? 

We conducted a two-fold research study. The first study explored malicious selling strategies during livestream shopping and how platform designs mediated these strategies, through a qualitative analysis of 40 livestream shopping sessions from two of the top livestreaming platforms in China, i.e., Taobao and TikTok. The second study sought to provide viewers' perspectives about malicious selling strategies through semi-structured interviews with 13 participants who had shopping experiences on livestreaming platforms in China. 

From the first study, we identified 16 malicious selling strategies and categorized them into four categories (i.e., Restrictive, Deceptive, Covert, and Asymmetric) that were defined by Mathur and colleagues~\cite{mathur2019dark}. Among the 16 malicious selling strategies, nine malicious strategies were amplified by platforms' designs. 
The interview study revealed that some participants were not aware of the malicious selling strategies or designs. For other participants, even though they were aware of malicious selling practises, they developed multiple ways to adapt to, or push back against, these strategies and designs. Participants also reported that they were faced with challenges when defending against malicious selling due to loosely-regulated and individualized streaming practices, and some system designs worsened the situation.  
Based on these findings, we discuss policy and design implications to counter malicious selling on livestream platforms.

The contributions of this work are two-fold. First, we identify 16 malicious selling strategies, some of which are new compared to existing typologies (e.g., Playacting). More than half of the strategies were amplified by platform designs, which led to negative effects on viewers when combined with streamer's malicious selling strategies.
Second, we provide a rich account of viewers' perspectives about how they perceived and reacted to streamers' malicious selling.

\section{Related Work}
This work aims to investigate streamers' malicious selling strategies and platform designs in livestream shopping applications. We first review studies about livestream shopping. Then, we review the selling strategies that have been investigated in traditional online shopping mediums (e.g., TV shopping). 
Last, we review literature specifically discussing the negative effects platform designs have on users in the online shopping and other scenarios.

\subsection{Livestreaming as an Emerging Online Shopping Form}
Livestreaming has been widely used as an engaging way to connect community members~\cite{hamilton2014streaming,lu2018you}.
Previous studies have explored engagement motivations, participation demographics, and engagement levels in various livestream communities, including livestreams for entertainment, education~\cite{chen2019integrating, hamilton2018collaborative}, skill improvement~\cite{lu2018streamwiki}, programming~\cite{faas2018watch}, cultural practices~\cite{lu2019feel}, creative activities~\cite{lu2018you}, and online communications~\cite{taylor2018watch}. 
Prior research has explored reasons for livestream participation from both streamer perspectives~\cite{kaytoue2012watch, lottridge2017third, pellicone2017game} and viewer perspectives~\cite{scheibe2016information, greenberg2016interaction, hamilton2014streaming}, how people formed communities between streamers and viewers via livestreams~\cite{hamilton2014path}, and different types of interactions within livestreaming communities~\cite{greenberg2016interaction, lessel2017expanding, pires2015youtube, zhu2017understanding}. 
For example, streamers on livestream platforms such as Twitch, broadcasted the playing of video games to share gameplay skills and common interests with viewers~\cite{hamilton2014streaming, cook2019behind, pellicone2017game, sher2019speedrunning, wohn2018explaining}.
Prior work has also analyzed viewers' negative behaviors during livestreams and how livestreaming platforms allowed viewers to report inappropriate content~\cite{wohn2019volunteer}. Livestream platforms have also integrated communication channels to increase viewer engagement~\cite{clickman2018design, lessel2017expanding}. 
This high degree of interactivity and engagement results in some viewers being more willing to purchase products created by streamers during a livestream~\cite{lu2019feel} or the merchandise designed by the streamers~\cite{sher2019speedrunning}. This suggests that there is a great potential for livestreaming to be a new medium for e-commerce.

Livestreaming e-commerce, or livestream shopping, has attracted researchers' attention in recent years. 
Some research has focused on viewers' (i.e., buyers) motivations: why they watch livestreams and buy products during them~\cite{cai2018utilitarian,xu2020drives,li2020understanding,cai2019live,su2019empirical}.
By analyzing self-reported data, scholars found that utilitarian and hedonic values were two of the main values that motivated customers to shop during livestreams. Utilitarian value was related to purchase needs, whereas hedonic value was related to affective commitments to streamers~\cite{cai2018utilitarian,li2020understanding}. 
Researchers have also found that trust was an intermediate variable that transferred affective commitment to engagement and buying on livestream shopping platforms~\cite{chen2020livestreaming}. 

A few researchers have also focused on the streamers' (i.e., sellers) perspectives. Researchers have reported a range of selling strategies that streamers used for increasing sales revenue~\cite{wongkitrungrueng2020live,chen2019everyone}. But few of them unpacked  how platform designs may mediate streamers' behaviors and how these streamer practices may impact viewer's buying experiences. This study aims to explore the questions. In the following, we review selling behaviors in the virtual context (e.g., TV shopping).

\subsection{Seller Behaviors on Online Shopping Platforms}


Although livestream shopping is emerging, it has similarities to \textit{TV shopping} and \textit{social shopping} given all the shopping modalities are happening virtually. To better understand seller behaviors and selling strategies on livestream shopping platforms, we review well-studied selling behaviors in TV shopping and social shopping.

TV shopping is a shopping method dating back to the 1980's and is still popular in many countries. During a pre-recorded TV shopping program, a TV host presents products in detail, and viewers can order the products by calling the telephone number shown on the screen~\cite{grant1991television}. Both livestream shopping and TV shopping show products in a lively way, however livestream shopping is more interactive than TV shopping~\cite{liu2021learn}. 
Many studies have investigated selling strategies in TV shopping. For example, Fritchie and colleagues found that social proof, scarcity, authority, commitment and consistency, liking, and reciprocation were used as selling strategies to simulate selling. These selling strategies were based on persuasion theory~\cite{cialdini2009influence}. TV hosts also built personal and emotional connections with viewers to fostered viewer trust ~\cite{kline2005interactive,stephens1996enhancing}. For example, hosts asked questions to establish similarities with viewers~\cite{stephens1996enhancing}.


Social shopping is another form of online shopping, by which product promotions are disseminated through personal social networks on online platforms (e.g., Facebook, Instagram) negating the need to switch to other e-commerce channels~\cite{chen2020understanding}. The salespeople, or ``market intermediaries,'' actively share product information with their social connections on social shopping platforms. Some research has studied the behaviors of market intermediaries in social shopping platforms and found that market intermediaries helped customers detect trending products, evaluated products, and provided a good price and discount to customers~\cite{chen2020understanding,chen2002referral,kim2013effects}. In addition to providing product-related support, they also built trust with potential customers by generating and sharing content (e.g., short video)~\cite{hajli2015social} and providing emotional support~\cite{shanmugam2016applications}. Customers reported that, in addition to trust, culture and social presence were two other factors that impacted market intermediaries' sales~\cite{ng2013intention}.

Among these selling strategies, researchers have also examined the malicious side of selling. The existing literature has extensively discussed the negative side of TV selling strategies. For example, consumers believed TV shopping was extremely risky~\cite{burgess2003comparison} as the product quality might be poor and the price was uncertain~\cite{burgess1995television}. 
TV shopping hosts were also found to use their knowledge of human behaviors and viewers' desires to implement deceptive strategies that incited purchasing. For example, TV shopping used compliance-gaining techniques to urge viewers. These techniques included using countdown timers and ``sold out'' signs~\cite{auter1993buying}. 
The host of TV shopping were also found to introduce a product with emphasis on financial prudence (e.g., bargain price and versatility) and upward mobility (e.g., items looking expensive)~\cite{cook2000consumer}. Studies also revealed undisclosed advertising strategies in social shopping~\cite{wu2016youtube,mathur2018endorsements}. Undisclosed advertising is a form of advertising where influencers on social media platforms such as YouTube do not explicitly disclose their collaborative relationships with advertisers or products. Undisclosed advertising is prevalent on social media platforms, but it is hard for users to be aware of ambiguous disclosures~\cite{mathur2018endorsements}.

Livestream shopping combines salesmanship and parasocial interaction characteristics. During a livestream, a streamer serves as the salesman to display products and build social connections with viewers. For example, a streamer who sells books in her channel may also teach history courses relevant to the promoted books and share her personal stories during her livestream.

On livestream shopping platforms, streamers can use various strategies to attract attention, increase sales, and build customer engagement. For example, Wongkitrungrueng et al.~\cite{wongkitrungrueng2020live} found four sales approaches were used in livestream commerce, i.e., ``transaction-based,'' ``persuasion-based,'' ``content-based,'' and ``relationship-based.'' 
Through the transaction-based approach, streamers used simple selling, limit quantity/time/offer, and demonstration to facilitate their selling. Through the persuasion-based approach, streamers relied on game-prize, show, and personal characters to persuade viewers. Through the content-based approach, sellers provided product related or non-product related information to enrich their livestream's content. 
Through the relationship-based approach, streamers shared their personal life, feelings or experiences, and organized community activities to strengthen their relationships with viewers.

These characteristics result in livestream shopping differing from TV shopping and social shopping. It signals a need to investigate whether malicious selling strategies are also existing in livestream shopping. To our best knowledge, the only work that studied malicious selling in livestream is~\cite{xu2019investigation}. The study suggested that a viewer's impulsive buying behavior might be related to ``emotional energy,'' which refers to customers' affection with streamers. In this work, we aim to understand what streamers' malicious selling strategies in livestream shopping are and how livestream platform designs mediate malicious selling strategies. To address the latter question, we review the scholarship on the negative effects of interface designs on users.

\subsection{Online Shopping Platform Design and Deceptive Designs}
Livestream shopping is a new type of online shopping experience, thus streamers' and viewers' experience are empowered or bounded by a platform's UI designs. In the HCI community, one avenue of research that focuses on the negative effects of interface designs on users is ``dark patterns.'' As a concept referring to marketing deceptive practices, nudging public policy and others, dark pattern does not yet have a consistent definition. Brignull et al. initially defined it as ``a user interface that has been carefully crafted to trick users into doing things''~\cite{brignull2015dark}. Later, Brignull used a new term ``deceptive design'' to emphasize the deceptive attribute in the definition and make the term more inclusive~\cite{brignull2018dark}. We bring this change to the front and use the new term in the following\footnote{We follow the ACM's inclusive terminology policy \url{https://www.acm.org/diversity-inclusion/words-matter} and transition to the new term ``deceptive design'' in this work regardless of what prior literature used. We also urge researchers to use the new term in future work.}. Narayanan et al. then added the covert manipulation and coercion aspects and defined that deceptive designs ``deceive users while others covertly manipulate or coerce them into choices that are not in their best interests''~\cite{narayanan2020dark}. These definitions emphasize that deceptive designs undermine individual decision-making and harm their interests~\cite{mathur2021makes}. 

Researchers have developed taxonomies to classify deceptive designs. Brignull proposed a deceptive design taxonomy to categorize the various designs~\cite{brignull2015dark,brignull2018dark}. The taxonomy consists of 12 categories, e.g., ``Bait and Switch,'' ``Disguised Ad''
~\cite{brignull2018dark}.
Gray and colleagues criticized that the taxonomy mixed up context, strategy, and outcome. Thus they simplified the taxonomy into five categories: ``nagging,'' ``obstruction,'' ``sneaking,'' ``interface interference,'' and ``forced action''~\cite{gray2018dark}.
Mathur and colleagues built upon this research and developed a deceptive design taxonomy specifically for online shopping websites~\cite{mathur2019dark}. The taxonomy included seven deceptive design categories. Compared to Bringnull's and Gray's taxonomies~\cite{gray2018dark}, Mathur's version added ``Urgency,'' ``Social Proof,'' and ``Scarcity'' for online shopping context, and they combined multiple categories such as ``Confirmshaming'' and ``Trick Questions'' into a single ``Misdirection'' category. Further, Mathur and colleagues summarized five characteristics of deceptive designs, ``Asymmetric,'' ``Covert,'' ``Deceptive,'' ``Hides Information,'' and ``Restrictive''~\cite{mathur2019dark}. Deceptive designs need to meet at least one of the five characteristics. Gray et al. stated that deceptive designs should be ``obnoxious, coercive, or deceitful''~\cite{gray2020kind}. These properties serve to differentiate deceptive designs from non-deceptive designs.
Deceptive designs have been reported in various scenarios including e-shopping websites~\cite{mathur2019dark,moser2019impulse}, mobile applications~\cite{di2020ui}, gaming~\cite{deterding2020against,zagal2013dark}, robot~\cite{lacey2019cuteness}, etc. Studies that were not particularly looking at deceptive designs also reported ``bad'' design practices that could potentially harm users, such as malicious interface techniques~\cite{conti2010malicious}, asshole designers~\cite{gray2020kind}, privacy paradoxes~\cite{waldman2020cognitive}, and impulsive buying~\cite{moser2019impulse}. 
Conti and Sobiesk defined malicious interface techniques as the intention to ``deliberately sacrifice the user experience'' to achieve designers' goals~\cite{conti2010malicious}.
Gray et al. found that even without the presence of deceptive designs, asshole designs could still deceive users, such as restricting users' tasks, setting traps that were hard to avoid, and misrepresenting information, which highlighted designers' unethical intentions and motivations~\cite{gray2020kind}. 
Moser and colleagues~\cite{moser2019impulse} studied impulse buying in e-shopping websites. Some impulsive buying features were different from deceptive designs. For example, an impulse buying feature ``Offering Returns'' is not a deceptive design because it does not directly benefit e-shopping websites~\cite{brignull2015dark,mathur2019dark}.

Studies have started to understand users' perceptions about deceptive designs, i.e., whether users can be aware of deceptive designs~\cite{di2020ui,maier2020dark,waldman2020cognitive} and how deceptive designs influence user behavior~\cite{luguri2019shining}. Studies indicated that users tended to be insensitive to deceptive designs even with hints~\cite{di2020ui,maier2020dark}. For example, Di Geronimo and colleagues explored user's perception of UI deceptive designs in mobile applications. Their online user experiment illustrated that the presence of malicious designs was not easy for users to notice~\cite{di2020ui}. If users were informed of the existence of deceptive designs, however, they were found to be better aware of them~\cite{waldman2020cognitive}.
Maier and Harr~\cite{maier2020dark} found that deceptive techniques were perceived as less malicious by users and some of them were regarded as just sneaky and dishonest. 
Luguri and Strahilevitz found that well-educated users were less susceptible to deceptive designs than a less-educated group. The deceptive designs ``hidden information,'' ``trick questions,'' and ``obstruction'' were more likely to be manipulative than others such as ``nagging,'' ``sneaking,'' ``forced action''~\cite{luguri2019shining}.
Mathur et al. discussed the origins of deceptive designs from the disciplines of psychology and behavioral economics and proposed possibilities to evaluate the impacts of deceptive designs on individuals such as financial loss, violation of privacy and cognitive burden~\cite{mathur2021makes}. These findings indicate the need to better understand users' perceptions of, and reactions to, deceptive designs with the purpose of mitigating harm. 

In this present research, with the reported malicious selling in livestream shopping, there may be some designs that impair viewers' decision-making and purchase behaviors due to the ways that streamers use these designs for malicious selling purposes. 
Thus, we investigated what malicious selling strategies were used by streamers on livestream shopping platforms and how these designs might fluidly turn into deceptive designs during livestream shopping scenarios.





\section{Methods}
A two-phase study was conducted to address the research study. The first phase sought to identify malicious selling strategies and platform designs in recorded livestream sessions via qualitative content analysis (i.e., RQ1\& RQ2). In the second phase, an interview study triangulated the findings from the first phase from viewers' perspectives (i.e., RQ3).

\subsection{Qualitative Analysis of Livestream Shopping Video Recordings}
To investigate streamer selling strategies, a group of 40 publicly available livestream sessions were sampled and archived as video recordings.
These 40 sessions were taken from two predominant livestream shopping platforms in China, Alibaba's Taobao and ByteDance's TikTok.
To identify malicious selling strategies, inductive coding was performed with the recorded videos. 

\subsubsection{Using Two Livestream Platforms as Research Sites}
\begin{figure}
  \subfloat[A TikTok Livestream Channel.]{%
	\begin{minipage}[c][1.3\width]{
	   0.49\textwidth}
	   \centering
	   \includegraphics[width=0.9\textwidth]{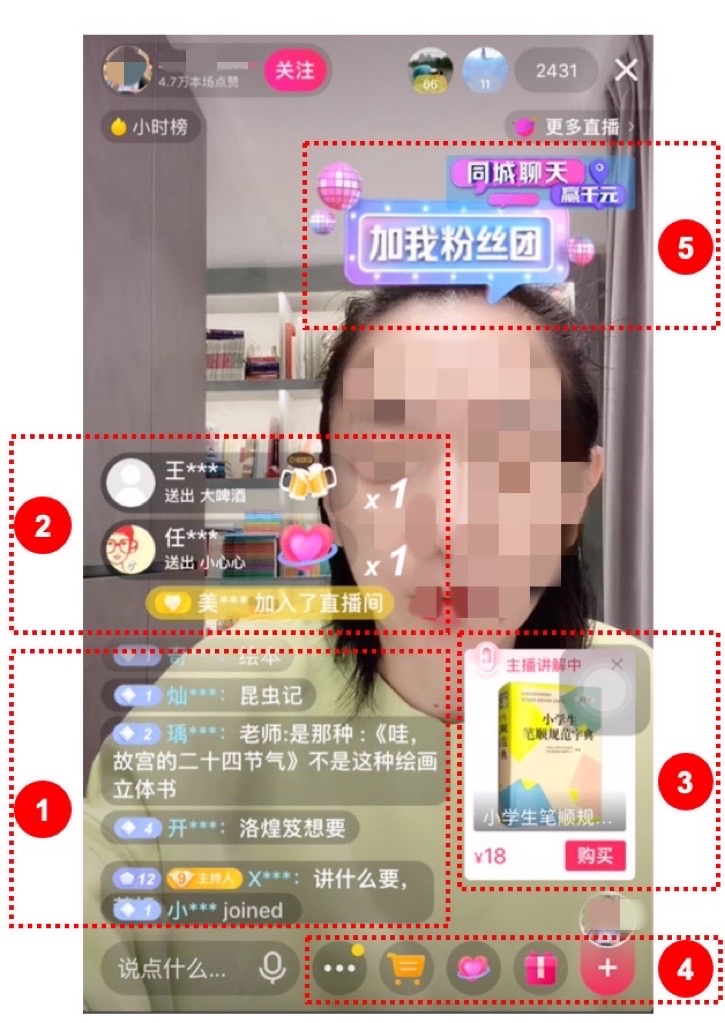}%
	   \label{fig:douyin}%
	\end{minipage}}%
 \hfill	
  \subfloat[The Shopping Cart on TikTok.]{%
	\begin{minipage}[c][1.3\width]{
	   0.49\textwidth}
	   \centering
	   \includegraphics[width=0.9\textwidth]{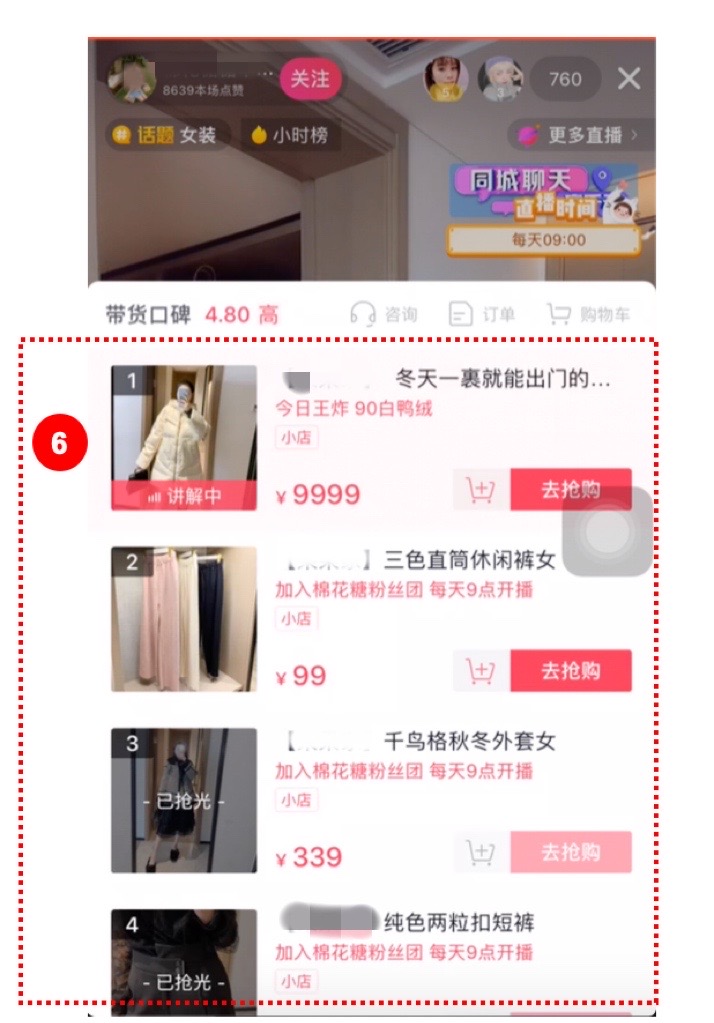}%
	   \label{fig:shopcart}%
	\end{minipage}}%
\caption{The TikTok interface of \protect\subref{fig:douyin} a livestream channel and \protect\subref{fig:shopcart} shopping cart on. The main features are: 1) a chat channel where both viewers and the streamer's assistant can post comments; 2) from the top down, viewers' action messages, including a sending gifts (to the streamer) message (\textit{``User A (王***) is sending beer''}), a sending likes message (\textit{``User B (任***) is sending likes''}), a joining the livestream channel message (\textit{``User C (美***) is joining the channel''}); 3) the link of the streamed product; 4) other function buttons:  from left to right, more function button, checking shopping cart button, sending likes button, sending gifts button, and checking the streamer's profile button; 5) the subscribing reminder tag, named ``Join my fans group''; 6) the shopping cart with numbered products, one ``Streaming'' mark on the 1st product, and two ``Sold out'' marks on the 3rd and 4th products.}
\label{fig:interface}
\end{figure}

Two top livestream commerce platforms, Alibaba's Taobao and ByteDance's TikTok, were selected for analysis. They are the top two livestream shopping platforms in China~\cite{traction2020}. 
Taobao is originally an online shopping platform, thus its livestream feature is an extension to its online shopping website. 
TikTok is a short video sharing and livesteam platform, thus their add-on shopping feature is an extension to its livestream service~\cite{kpmg2020}. \autoref{fig:interface} demonstrates the primary features on the TikTok livestream platform. 

Due to their different origins, the livestream styles and streamer backgrounds were slightly different on these two platforms.
The streamers on the Taobao livestream platform were small business owners or temporary streamers (e.g., salesmen) hired by business owners and they sold products from their own business. 
In contrast, the streamers on TikTok livestream shopping were  celebrities, influencers, or professional streamers who collaborated with brands and business owners. 
Functionality-wise, Taobao livestreams redirected viewers to an online store on the Taobao website;
while TikTok provided a ``shopping window'' feature within the stream session, enabling users to browse a list of products and complete purchases with one click without leaving the livestream session~\cite{Shopwindow}.

\subsubsection{Data Collection of Livestream Shopping Sessions}
Streamers were sampled based on the number of subscriptions, the product type, and livestreaming time of the day.
Some streamers had a huge subscription population (e.g., two top streamers on Taobao had millions of subscribers) whereas some others had a relatively small subscription population (e.g., a farmer streamer had only hundreds of subscribers). To draw a diverse sample, we selected both streamers with small and large numbers of subscribers.
We also sampled accounts to cover diverse product types. On Taobao livestream platform, stream channels were categorized by the type of products, i.e., Food, Apparel, Beauty, Jewelry, Mother\&Kid, Home, and Electric. On TikTok, there were no specific livestream shopping categories. 
Thus, we sampled streamers' accounts on the two platforms based on Taobao's seven product categories. While most livestream sessions occurred in the evening to gain higher customer volume, there were  livestream sessions occurring in the morning and afternoon. Thus, we selected three time segments for data collection: the morning segment (10 AM - 12 PM China Standard Time), the afternoon segment (1 PM - 3 PM), and the evening segment (8 PM - 11 PM). The evening segment was longer than the other two segments to accommodate for the increased number of livestreaming sessions occurring in the evening.  
The first author conducted the recording task to make sure the sampling strategy was consistent. The researcher used screen recording software to save livestream sessions as video recordings. The researcher used an incognito mode to log on to Taobao and TikTok to avoid any recommendation algorithm bias. All livestream sessions were publicly available.

Normally, one livestream session ran for a few hours each day. During the session, a streamer displayed multiple products in turn. To conduct a comprehensive investigation of malicious selling strategies during livestream shopping, we needed to cover diverse selling behaviors. Therefore, there was a trade-off between data diversity and data size. To decide the length of livestream sessions to record, we first recorded five livestream sessions from each platform. Each recording lasted for 15 to 25 minutes. We found that streamers usually spent about five minutes on a single product, and streamers' selling behaviors did not change much from product to product. 
Thus, we decided to shorten the length of recordings to approximately seven minutes. We collected another 15 livestream sessions from each platform. In total, we collected 40 recordings, 20 from Taobao and 20 from TikTok \autoref{tab:recording}. 

\begin{table}[]
\centering
\caption{The 40 sampled livesteam sessions from Taobao and TikTok, including the video ID, the livestream channel's name in Chinese, the number of subscriptions, the number of real-time viewers, the category of streamed shopping products. Note that the product category ``All'' indicates that the livestream channel sold multiple types of products.}
\label{tab:recording}
\resizebox{\textwidth}{!}{%
\begin{tabular}{llllllllll}\toprule
\textbf{Video ID} &
\textbf{Plateform} &
  \textbf{Channel Name} &
  \textbf{\# of Subscribers} &
  \textbf{\# of Viewers} &
  \textbf{Product Category}
  \\\midrule
VT1 & Taobao & 薇***a      & 39M  & 1.9M & All             \\
VT2 & Taobao & 李***琦         & 38M  &  1.3M    & All         \\
VT3 & Taobao & v***d     & 14M  & 11K  & Apparel         \\
VT4 & Taobao & C***师       & 898K & 13K  & Beauty       \\
VT5 & Taobao & 小***a       & 776K & 657K & Apparel         \\
VT6 & Taobao & 李***响          & 537K & 613K & All           \\
VT7  & Taobao & 美***器      & 348K & 461  & Electronics  \\
VT8 & Taobao & 老***珀      & 178K & 39K  & Jewelry      \\
VT9  & Taobao& 淘***装     & 146K & 334  & Mother \& Kid &      \\
VT10 & Taobao & 混***赵 & 83K  & 869  & Beauty        \\
VT11  & Taobao    & 栗***品    & 51K        & 5.0K          & Mother \& Kid  \\
VT12 & Taobao & 南***大       & 27K  & 6.0K & Food         \\
VT13 & Taobao & 葡***玩     & 27K  & 1.7K & Mother \& Kid     \\
VT14  & Taobao & 奢***A     & 21K  & 741  & Apparel      \\
VT15 & Taobao & 新***昕     & 18K  & 4.7K & Jewelry        \\
VT16  & Taobao & 冰***y    & 9.5K & 33K  & Beauty    \\
VT17 & Taobao & 小***直     & 7.5K & 5.8K & Apparel      \\
VT18 & Taobao & 石***道         & 3.5K & 1.5K & Jewelry      \\
VT19 & Taobao & 罗***点     & 2.3K & 23K  & Home            \\
VT20 & Taobao & 李***店      & 976  & 21K  & Food            \\[3pt]
VD1 & TikTok     & 主***芳       & 7.7M       & 2.4K          & Education\\
VD2 & TikTok & 恩***克           & 4.7M & 2.6K & Food\\
VD3 & TikTok & 养***明        & 4.7M & 2.4K & Food\\
VD4 & TikTok & 晨***强     & 4.0M & 7.0K & Beauty\\
VD5 & TikTok & 三***鼠         & 3.3M & 4.0K & Food\\
VD6 & TikTok & 彩***妇         & 2.9M & 4.6K & All\\
VD7 & TikTok & 贾***坊        & 1.9M & 2.9K & Apparel\\
VD8 & TikTok & 圆***到         & 1.6M & 3.4K & Home\\
VD9 & TikTok & 新***厂     & 1.4M & 3.9K & Apparel\\
VD10 & TikTok & 新***店       & 1.2M & 1.6K & Apparel\\
VD11 & TikTok & 棉***播 & 795K & 730  & Apparel\\
VD12 & TikTok & 丹***鲜        & 777K & 1.7K & Food\\
VD13 & TikTok & 美***）    & 532K & 3.4K & Jewelry\\
VD14 & TikTok & 阿***炸         & 412K & 1.7K & All\\
VD15 & TikTok & 每***居         & 387K & 1.7K & Home\\
VD16 & TikTok & 易***号      & 374K & 2.9K & Apparel\\
VD17 & TikTok & 自***堂          & 368K & 50   & Beauty\\
VD18 & TikTok & U***制         & 70K  & 1.4K & Apparel\\
VD19 & TikTok & 祥***丸      & 45K  & 5.7K & Food\\
VD20 & TikTok & 湘***菜      & 26K  & 211  & Food\\\bottomrule   
\end{tabular}%
}
\end{table}

\subsubsection{Qualitative Coding Procedure}

There was an overlap between our coding and the existing typologies~\cite{brignull2015dark,brignull2018dark,mathur2019dark,gray2018dark} so we first conducted a deductive analysis on the first 10 recordings (i.e., five from Taobao and five from TikTok) using codes from the existing typologies. The coding was primarily focused on streamers' malicious selling strategies and designs used by streamers together with such strategies. Each researcher viewed the first ten recordings independently and coded multiple video elements, including streamers' speeches and actions, the features or functions used by streamers, the content of comments in the chat channel, the products sold, etc. We wrote down the codes along with the timestamps in an Excel sheet and took notes. All researchers had prior shopping experiences on China livestream platforms.

Notably, we defined malicious selling as ``streamers leveraging unbalanced power and information to deceive, coerce, or manipulate viewers into buying that was against viewers' best interests.'' We carefully checked if the selling strategies met the definition. Since livestreams usually do not allow for post stream checking\footnote{Livestream channels rarely record their videos and allow viewers or subscribers to view them again.}, we recorded evidence during the data collection for post check if there would be the potential for a malicious selling strategy. For example, if a streamer applied the ``urgency'' selling strategy for a product or a discount, we checked whether the product or the discount was expired when the streamer said it would. If the product or the discount was still available, then the ``urgency'' information mentioned by the streamer was fake. We recorded the check process in the videos.

We discussed the deductive coding results together and refined the codebook. We recognized that new codes were derived from the data. Then, we performed inductive coding. Specifically, we compared the new codes with the existing typologies and added the new codes to the codebook, for example, ``Playacting'' (engaging in pretense to induce viewers to buy or to gain affection and sympathy) and ``Sophistry'' (defending for self with a deliberately invalid argument). We then coded another 10 recordings using the updated codebook. No new codes emerged during the process. The codebook reached saturation. An inter-coder reliability~\cite{krippendorff2004reliability} of 0.91 was computed (Cohen's kappa). Using the codebook, we then coded the remaining 20 recordings. 

In total, we identified 15 malicious selling strategies from the 40 recording sessions (\autoref{tab:strategies}). We borrowed or referred to codes such as ``urgency,'' ``scarcity,'' ``social proof,'' ``forced action,'' and ``nagging'' from existing typologies~\cite{mathur2019dark,gray2018dark,crabtree1992doing}. We added new codes such as ``Playacting'' and ``Sophistry'' to our taxonomy. The new codes were marked in bold in~\autoref{tab:strategies}. We found that these strategies fit into characteristics of deceptive designs initially defined by~\cite{mathur2019dark}. Thus, we further grouped the malicious selling strategies into four categories: Restrictive, Deceptive, Covert, and Asymmetric. We re-defined the four categories in the livestream shopping context in the Results section. 

During the coding, we also identified 15 interface designs that mediated the malicious selling strategies \autoref{tab:designs}. These designs are also explained in more detail in the Results section. 

\subsection{Interview Study}
To understand viewer perceptions of these malicious selling strategies, we designed and conducted a semi-structured interview study to gather viewers' awareness, attitudes, and reactions to these selling strategies. This study was approved by the IRB of the fourth author's university.

\subsubsection{Participant Recruitment}
Participants who had purchase history on livestream platforms in China were recruited to participate using snowball sampling. Recruitment flyers were disseminated through researchers' social networks and the research team also asked participants to spread the recruitment flyers to their acquaintances. Once one agreed to participate in the interview, they were asked to complete a screening survey. This survey included questions such as ``Have you ever watched livestream for shopping?'', ``How frequently do you watch livestream?'', ``Have you ever bought products on livestream?'', ``What platform(s) do you watch livestream?'', and so on.

Twenty-three responses to this survey were received. We filtered out the respondents who did not have purchase history on livestream platforms or did not provide contact information and interviewed participants from the remaining responses. We kept tracking the insights reported by participants while interviewing participants and found no more new insights reported by P11, P12, and P13. Thus, we stopped recruiting participants. In total, we interviewed 13 participates. \autoref{tab:participants} presents the participants' demographic information. Out of the 13 participants, 11 participants were female. All participants had a high school degree or higher and were from cities or towns in China. All had online shopping experiences either from Taobao or TikTok or both before the interview. 
According to a China livestream e-commerce user report in 2020, the ratio of male users to female users on Taobao and TikTok livestream commerce platforms was about 3:7, and users aged between 18 and 40 was about 85\%~\cite{oratings2020live}. Thus, our participant sample represented the user population on the two platforms with a slight skew towards female.

\begin{table*}[]
\centering
\caption{The summary of participants' demographic information (i.e., age and gender) and general experiences watching livestream for shopping (i.e., how often they watched livestreams, how long they had watched livestreams, and on what platforms they watched livestreams). }
\label{tab:participants}
\resizebox{\textwidth}{!}{%
\begin{tabular}{llllll}
\toprule
\textbf{Participant} &
  \textbf{Age} &
  \textbf{Gender} &
  \textbf{Frequency of Viewing } &
  \textbf{History of Viewing} &
  \textbf{Platform(s)} \\\midrule
P01 & 26-35 & Female  & Once a week          & 1-3 months           & Taobao, TikTok         \\
P02 & 26-35 & Male           & Once a week          & 4-6 months           & Taobao, TikTok         \\
P03 & 18-25 & Female           & Several times a week & 7-12 months          & Taobao, TikTok         \\
P04 & 18-25 & Female        & Once a week          & 7-12 months          & Taobao                 \\
P05 & 26-35 & Female                           & Several times a week & 1-2 years            & Taobao, TikTok         \\
P06 & 26-35 & Female                             & Every day            & 4-6 months           & Taobao, TikTok, WeChat \\
P07 & 18-25 & Female                           & Several times a week & 7-12 months          & Taobao                 \\
P08 & 36-45 & Male    & Once a week          & 1-2 years            & Taobao                 \\
P09 & 26-35 & Female      & Several times a week & 1-2 years            & Taobao                 \\
P10 & 36-45 & Female                          & Once a week          & 4-6 months           & Taobao, TikTok         \\
P11 & 36-45 & Female                            & Several times a week & \textgreater 2 years & Taobao, TikTok         \\
P12 & 36-45 & Female                          & Several times a week & 7-12 months          & Taobao                 \\
P13 & 18-25 & Female          & Once a week          & 1-2 years            & Taobao, TikTok  \\\bottomrule      
\end{tabular}%
}
\end{table*}

\subsubsection{Interview Procedure}
The interviews were conducted from December 2020 to January 2021 and used video or audio conferencing tools (i.e., WeChat, Zoom) to interview participants based on their preferences. All interviews were conducted in Mandarin Chinese. The first author led the interviews, asked questions, and recorded the audio. The second author took notes. Both researchers were fluent in Mandarin Chinese. Before conducting the interviews, pilot studies were run with two volunteers to test the interview procedure. All researchers participated in the pilot studies and provided feedback. The pilot results were excluded from the results. 

The interview contained two primary sections: the first section investigated what malicious selling strategies participants experienced, and the second section let participants share their experiences about how they reacted to malicious selling. 

At the beginning of the interview, we started with some icebreaker questions, such as ``How often do you watch livestreams?'' or ``How long have you watched livestreams for shopping?'' Then, we dove into participants' shopping experiences on livestream platforms. We asked ``Have you bought something on any livestream platforms?'' Then, we asked participants to talk about their most recent/impressive shopping experience in detail.
We also asked the participants about their perceptions of streamers and other viewers in the livestream channel(s). 

The questions were designed to help participants recall as many of their shopping experiences in livestream channels as possible. If the participant had difficulty recalling an experience, we asked them to show us the livestream platform and describe the interface design. This technique helped participants recall their, and streamers', behaviors. 

To avoid leading questions that could bias participants' perceived malicious selling strategies, we did not remind participants of any specific malicious strategy. Instead, when they mentioned experiences related to the strategies in our taxonomy, we asked follow-up questions regarding participants' attitudes, such as ``Do you trust the streamer?'' or ``Are you satisfied with the product?'' 
Then, we probed to further explain why they did or did not think so and how they reacted to this. 
If participants never mentioned any malicious selling strategy, we asked ``Have you noticed anything inappropriate or wrong, or that you feel uncomfortable with, including the streamer, the streamer's assistant, viewers, and the interface designs?'' If participants answered no, we did not ask any probing questions.
At the end, participants were asked to suggest ways to improve the livestream shopping experience. 

Each interview lasted approximately 40-50 minutes. Participation in the interview was voluntary. All interviews were audio-recorded with participants' approval and transcribed to text. 

\subsubsection{Data Analysis}
The first and second authors conducted an open coding process derived from the grounded theory method~\cite{strauss1994grounded}. They coded the transcripts at the sentence-level independently. 
All researchers discussed the coding results together and grouped the codes into six themes (i.e.,  general experiences, motivations for watching livestreams for shopping, purchase procedures, awareness of malicious selling strategies, ways of coping with malicious selling strategies, and challenges when defending against malicious strategies). 
Then, the researchers applied an axial coding method~\cite{creswell2016qualitative} and generated the final coding schema with three primary categories: awareness and attitudes towards malicious selling strategies, ways of coping with malicious selling, and challenges of coping with malicious selling.

\subsection{Ethics Discussion}
First, as the purpose of this research was not to identify individual livestream sellers who were applying malicious selling strategies, we de-identified the streamers' channel names. Note that most of the studied streamers' selling strategies and tactics were not malicious. Readers should not be misled or biased into thinking negatively about all platforms or steamers. To avoid general marketing tactics that mistakenly were identified as malicious selling strategies, the researcher team conducted several rounds of discussion to refine the definitions of malicious selling strategies in the taxonomy.

Moreover, a critical perspective was used to analyze livestream selling behaviors. We think it is critical to pay attention to any selling strategies or livestream design that undermine viewers' decision-making. Thus, we reported all livestream selling strategies either identified by researchers or participants, no matter whether these strategies are legal or prevalent.

\section{Results}
This section presents the results from the analysis of video recordings and the interview study. Then, we present the interview findings relating to how participants overcame malicious selling techniques and what their challenges were.

\subsection{Malicious Selling Strategies on Livestream Shopping Platforms (RQ1)}

The content analysis of the livestream shopping video recordings revealed 15 malicious selling strategies. Participants were aware of most malicious strategies except for the Pressured Selling strategy.
Moreover, three participants mentioned a new malicious selling strategy, Forced Wholesale. Notably, if a participant mentioned a livestream shopping experience related to a malicious selling strategy but did not perceive it as malicious, we did not count the participant for the malicious strategy. We categorized the 16 malicious selling strategies into four types---Restrictive, Deceptive, Covert, and Asymmetric---which were developed by~\cite{mathur2019dark}.

\autoref{tab:strategies} lists the four categories with 16 malicious selling strategies, the explanations, and the participants who were aware of the malicious strategies.

\begin{table*}
\small
\centering
\caption{The taxonomy of malicious selling strategies in livestream shopping: four categories, 16 strategies, the descriptions, and whether they were perceived by the participants.}
\label{tab:strategies}
\resizebox{\textwidth}{!}{%
\begin{tabular}{p{0.11\linewidth}p{0.23\linewidth}p{0.52\linewidth}p{0.14\linewidth}}
\toprule
\textbf{Category} &\textbf{Strategy} &
 \textbf{Description} & \textbf{Perceived By} \\\midrule
\multirow{3}{*}{Restrictive}&Forced Subscription &
 Forcing viewers to subscribe to the streamer and the streamer's fans group for rewards & P1, P2, P4\\[10pt]\cline{2-4}
& \textbf{Forced Endorsement} & Forcing viewers to send gifts to the streamer or sharing the streamer with others for rewards & P1, P7\\[10pt]\cline{2-4}
& \textbf{Forced Wholesale*} & Forcing viewers to buy multiple packs or units of a product &  P4, P7, P13\\\midrule
\multirow{4}{*}{Deceptive}&Fake Scarcity & Deceiving viewers about the high demand and limited availability of a product & P1, P3, P6, P11\\[10pt]\cline{2-4}
&Fake Urgency &
 Deceiving viewers that the product being sold or discount will expire soon & P2, P7, P9, P11 \\[10pt]\cline{2-4}
&Fake Social Proof &
 Highlighting fake positive comments, gifts and likes sent by astroturfers, or providing fake testimonials & P1, P2, P7, P8, P13\\[10pt]\cline{2-4}
& Fake Exclusive Pricing & Providing a fake, exclusive price or discount, or raising the original price before the  discount & P2, P9, P10, P13\\[10pt]\midrule
\multirow{3}{*}{Covert}& Visual Misrepresentation & Misrepresenting a product through visually beautifying or hiding aspects of it & P8, P10
  \\[10pt]\cline{2-4}
& \textbf{Playacting} &
 Engaging in pretend behaviors to induce viewers to buy or to gain affection and sympathy
  & P3, P7, P8, P11, P12\\[10pt]\cline{2-4}
&\textbf{Fuzzy Targeting} &
  Describing a product is applicable to all customer groups when it is not & P7, P8, P13
  \\[10pt]\midrule
\multirow{6}{*}{Asymmetric} & \textbf{Sophistry} &
 Defending oneself against negative comments using implausible language
  & P5\\[10pt]\cline{2-4}
  & \textbf{Retaining Customers} &
 Keeping viewers staying in the channel by holding unpredictable activities  & P7\\[10pt]\cline{2-4}
&\textbf{Egoistic Norms} &
 Asking viewers to follow  self-interested norms, and belittling or abusing viewers for anti-norm behaviors & P8\\[10pt]\cline{2-4}
&\textbf{Disgracing Others} &
Depreciating the quality of competitive products or the reputation of competitive streamers 
 & P5, P13 \\[10pt]\cline{2-4}
&Pressured Selling &
 Steering viewers into purchasing a more expensive version of a product or additional related products
 & / \\[10pt]\cline{2-4}
&Nagging &
 Repeatedly asking viewers to purchase without providing useful product related information & P10\\\bottomrule
\multicolumn{4}{l}{Strategies in bold were identified as new within this work.}\\
\multicolumn{4}{l}{*: Forced Wholesale is a malicious strategy that was only identified during the interview study.}
\end{tabular}%
}
\end{table*}

\subsubsection{\textbf{Restrictive Strategies}}
Restrictive strategies refer to those strategies where streamers asked viewers to take forced actions to obtain special rewards or prices. The Restrictive category contains three malicious selling strategies: Forced Subscription, Forced Endorsement, and Forced Wholesale.

\textbf{Forced Subscription} refers to when the streamer requires viewers to subscribe to the streamer's channel or the streamer's fans group for rewards (e.g., a discount, cashback). In this case, the subscription task is forced. In several channels, if viewers did not subscribe, they could not buy products at a discount.

\textbf{Forced Endorsement} is when a streamer asks viewers to send gifts, which need to be paid by viewers, or share the streamer's channel with others to obtain rewards (e.g., a discount). The endorsement task is also forced. Based on streamers' remarks during livestreams, they claimed that they could check whether viewers had subscribed to them or if they had performed endorsement tasks and then decide whether viewers could get a discount.

\textbf{Forced Wholesale} occurs when a streamer coerces viewers into buying multiple packs or units of a product in each order. Viewers had no choice to buy a single pack of a product. For example, P4 described a forced wholesale experience and commented that it was a common malicious strategy: ``I'll regret (to buy too many things). Streamers won't sell a single pack. It's kind of like wholesale. For example, if you want to buy toothpaste, you can only buy a set of four. It's a family package. If I buy it, I must regret since I can't use them out in a long time.'' (P4)

\subsubsection{\textbf{
Deceptive Strategies}}
Deceptive strategies refer to those wherein streamers used fake selling or product information to trick viewers to purchase products. The Deceptive category contains four selling strategies: Fake Scarcity, Fake Urgency, Fake Social Proof, and Fake Exclusive Pricing.

\textbf{Fake Scarcity} refers to streamers deceiving viewers about the high demand and limited availability of a product. Streamers used fake numbers to indicate the low stock of a certain product to make viewers believe that a limited number of items were available. It created an impression that viewers needed to make the purchase immediately. This strategy helps incite viewers' desirability for a product and increase the perceived value of the product~\cite{nodder2013evil,mittone2009scarcity}. 
For example, the streamer in VT19 warned that she had the last three blankets for viewers who would purchase in her channel. She said: ``You have to buy the bedclothes quickly to get one blanket, first come, first served.'' However, it turned out the blanket was a defaulted gift and everyone who purchased the bedclothes got a blanket for free. The channel displayed that more than three viewers bought the bedclothes and got the blankets but the streamer kept saying only three blankets were left. ``Three'' was a fake number to indicate the scarcity.

Streamers also claimed that their products were easy to be sold out in their livestream channels. The fast sold-out reminder induced viewers to buy. But no evidence showed that their products sold out fast after a while.

\textbf{Fake Urgency} occurred whenever a streamer tricked viewers into believing that a sale or discount would end soon or at a certain point in time in an attempt to speed up the viewer's decision to purchase. However, based on our analysis, the product selling or discount did not end as the streamer claimed. The Fake Urgency strategy was usually used in tandem with Fake Scarcity, taking advantage of viewers' scarcity bias and suggesting inaction would result in them losing potential savings~\cite{aggarwal2003use}. 

\textbf{Fake Social Proof} refers to when the streamer verbally highlighted fake positive information (e.g., positive comments, gifts, likes) sent by astroturfers, or provided fake testimonials for a product. Astroturfers (called ``water army'' in Chinese) were usually paid by streamers and pretended to send positive information in livestream channels. These messages created the impression that the product was very popular or of high quality. 

Streamers also claimed that their products had testimonials. However, some testimonials were just verbal claims and the testimonial origins were forged or unclear. For example, the streamer in VT12 claimed that her jeans were produced by a famous designer's factory. However, there was no label on the jeans or no testimonial presented on the product page to prove her claim. 

\textbf{Fake Exclusive Pricing} refers to the practises of providing a fake, exclusive price or discount or raising the original price before providing a discount. Streamers claimed that they had the lowest price or the best discount in their channels compared to anywhere else, which lacked evidence or was a lie. Other streamers raised their original prices. For example, one streamer raised the price of a set of pajamas to 189 RMB (about 30 U.S. dollars) before the livestream started. During the livestream, the streamer announced that he would sell the pajamas for 29.9 RMB (about five U.S. dollars), which was claimed to be an exclusive discount for viewers.

\subsubsection{Covert Strategies}
According to~\cite{mathur2019dark}, Covert means the effect or mechanism of deceptive designs is hidden from users, which create cognitive bias. But users can rarely realize it. One example is the decoy effect---providing an additional option to make other options more appealing~\cite{mathur2019dark}. Similarly, we define that Covert strategies make certain products' presence more appealing (than others) to influence viewers' decision-making; but their effect is covert, and users may fail to recognize the presence of these strategies. Covert selling strategies included Visual Misrepresentation, Playacting, and Fuzzy Targeting.

\textbf{Visual Misrepresentation} refers to situations where a streamer misrepresented a product by visually beatifying it product or hiding its defects. For example, P8 mentioned that one streamer used a type of special light in their livestream channel. The special light was different from natural light but made the products look better than they were: ``So the color of the stone in the livestream channel will look very good and very bright. But when you get it, the color of the stone is not quite the same as in the livestream channel. It is not very satisfactory, not the ideal effect.'' (P8)

\textbf{Playacting} is a pretend behavior used by streamers to induce viewers to buy or to gain affection and sympathy from viewers. The Playacting strategy leverages viewers' cognitive and emotional bias but viewers may not realize this strategy. For example, in VT5, when one product was sold out, the streamer begged her product manager to restock another 50 units of the same item. The product manager pretended that he tried so hard to find another 50 units for customers. However, this livestream channel sold the same item every day. The scarcity was not as claimed. The Playacting strategy was used for ``hunger marketing.'' Streamers also used playacting to fabricate a touching story or usage experience to emotionally stimulate viewers to place orders.

\textbf{Fuzzy Targeting} occurs whenever a streamer presents a product in a way to make it apply to all types of customers, regardless of their age, gender, or other characteristics. This approach expands the consumer pool by obscuring who are the target populations of a product. Viewers make the purchase without recognizing they were steered. For example, the streamer in VD20 recommended that her rice cake was a good snack for old people. Later she expanded the customer types to kids and adults. There were no specific targeted customers. The streamer applied the Fuzzy Targeting strategy when selling every product.

\subsubsection{Asymmetric Strategies} Asymmetric strategies refer to those where streamers ``impose unequal weights or burdens on the available choices presented to viewers''~\cite{mathur2019dark}. In the livestream shopping context, streamers created emotional or cognitive weights or burdens through verbal communication. The Asymmetric selling strategies identified were Sophistry, Retaining Customers, Egoistic Norms, Disgracing Others, Pressured Selling, and Nagging.

\textbf{Sophistry} refers to the streamer using implausible language to defend themselves while in an unfavorable situation during the livestream. The process of defending is granting the unfavorable situation more weight and imposing viewers to accept it. For example, in VT15, one viewer sent a message in the chat, saying that he was blocked after he contacted customer service for a wrong item shipment. The streamer was extremely angry when noticing this message in the public chat. To defend for herself, she said that ``You already got two bracelets from me, I bear the loss myself, why are you still mentioning the customer service in the chat?''

\textbf{Retaining Customers} refers to how viewers are more likely to buy or buy more products the longer the streamer keeps them in the channel. To keep viewers staying in their channel, streamers offer lotteries or make the sequence of promoted products unpredictable to viewers. For example, the streamer in VD1 promised to deliver a lecture at the beginning of the livestream. However, she kept selling her books throughout the livestream and did not start the lecture until the end of the livestream. The viewers could not predict when exactly the lecture started so they needed to watch the book selling and wait.

\textbf{Egoistic Norms} refers to when a streamer creates norms that are beneficial to themselves and asks viewers to follow the norms. The norms gain more weight and become more legitimate in the streamer's channel. If viewers do not follow the norms, they will be belittled or abused by the streamer for anti-norm behaviors. For example, in VD3, the streamer kept emphasizing that in his livestream channel, most viewers bought honey quickly without asking any questions. He said, ``If you (viewers) have any question about a product, you don't take me as your bro.''

\textbf{Disgracing Others} is a strategy where a streamer depreciates the quality of competitive products or the reputation of other competitive streamers. For example, the streamer in VD12 argued that some other competitors' (streamers') products were of low quality so that is why their prices were lower than hers. By disgracing others, the streamer was able to suggest that she had better quality products and made excuses for her higher prices.

\textbf{Pressured Selling} occurs when a streamer steers viewers into purchasing more expensive products or additional products. For example, in VD7, the streamer tricked viewers to buy the newer version of their shoes by listing the advantages of the newer version. However, the new version of the shoes was much more expensive than the older version, which attracted many viewers into the livestream session at the first place. 

\textbf{Nagging} refers to when a streamer repeatedly asks viewers to purchase products but at the same time, does not provide any useful information for viewers to make decisions about them. Non-malicious nagging existed in all video recordings. We only coded the malicious nagging strategy for streamers who pushed viewers to make orders without providing any useful production information.

Based on the recording coding and interview results, Restrictive, Deceptive, and Covert selling strategies were easier to perceive by participants. Asymmetric strategies were more difficult to perceive or verify. They were usually concealed and viewers were not aware that they were manipulated during the process. 
The qualitative analysis also revealed that malicious selling strategies were enhanced or hindered by platform designs. The next section presents how the four types of malicious selling strategies were mediated by livestreaming platform designs. 
\subsection{Platform Designs Mediating Malicious Selling Strategies (RQ2)}

\autoref{fig:interface} demonstrates the interfaces of a TikTok livestream channel. \autoref{tab:designs} shows 15 platform designs that could mediate the malicious selling strategies. The 15 designs were presented in both TikTok and Taobao platforms. The video recordings demonstrated that Restrictive selling strategies (e.g., Forced Endorsement, Forced Subscription) and Deceptive selling strategies (e.g., Fake Scarcity, Fake Social Proof, Fake Exclusive Pricing, and Fake Urgency) were more likely to be facilitated by platform designs than Covert and Asymmetric strategies. Covert and Asymmetric strategies relied more on sellers' sale skills or styles so they were less influenced by platform designs. Nine specific strategies were supported by design features or functions within the two livestream platforms. Below, we reflect on how livestreaming platform designs mediated the four categories of malicious selling strategies.

Designs using \textbf{subscription}, \textbf{reminder tags}, \textbf{sharing}, \textbf{liking}, \textbf{gifting} or \textbf{chat channels} enabled or supported Restrictive malicious selling strategies such as Forced Subscription and Forced Endorsement. It is common for e-shopping websites and livestream channels to apply these designs to require customers' or viewers' subscription, sharing, or gifting, however, we found that the backend filtering tool, which was orally claimed by streamers, was the key function to enable Restrictive malicious selling strategies. For example, before starting a lucky draw, the VT6 streamer told viewers that his assistant would check the viewers' subscription. If viewers did not subscribe to his channel, the streamer would cancel the viewers' prize. The VT7 streamer required viewers to click the ``like'' button and send a required message ``888'' (stands for fortune in Chinese pronunciation) to the public chat channel. She said: ``Everyone who wants a coupon sends 888.'' However, it is not clear whether and how the filtering tool ran. Viewers could only follow the streamer's requirement to obtain the coupon. Since Restrictive strategies were enabled or supported by platform designs, this suggests the need to modify or remove such design features to constrain malicious selling strategies.

The designs that enabled or supported Deceptive strategies presented misleading information (i.e., \textbf{notification messages}, \textbf{product pages}, \textbf{countdown timers}, \textbf{back-ordered warnings}, and \textbf{chat channels}) and inductive discounts (i.e., \textbf{red pockets}, \textbf{coupons}, or \textbf{flash sales}). These designs helped streamers create fake scarcity, social proof, exclusive pricing and urgency to induce viewers to buy products.
For example, both Taobao and TikTok have a notification message feature. Once a viewer adds an item to his/her shopping cart, there will be a public notification message on the screen saying ``User X is ordering the item''. TikTok also shows the total number of buyers on the screen. The public notification messages can be used as evidence of high demand. Using this evidence, streamers can urge viewers to place  an order. In one case, the streamer claimed, ``This is really the last one'', however, the streamer asked her assistant to restock the product later. Although Deceptive strategies were enabled or supported by specific designs, it took time for viewers to detect or verify fake information compared to the Restrictive strategies. It is thus necessary to provide tools to help viewers or customers to identify deceptive malicious selling.

Designs supporting \textbf{beautify camera filters} enabled and supported Covert strategies. For example, the visual misrepresentation enabled by the beautify camera filter facilitated malicious selling. The streamer of VD11 used the beautify camera filter provided by TikTok to hide the flaws of her apparel such as the bad texture and workmanship.


The results also revealed that the \textbf{chat channel} supported three types of malicious selling strategies. It was also the only design used by viewers to counter malicious selling strategies. This was not surprising because the chat channel was the only means for viewers to provide negative feedback. For example, some streamers enforced Egoistic Norms by asking viewers to send certain positive terms in the chat channel. Viewers sent comments such as ``poor quality products'' to fight against malicious selling via the chat channel.

\begin{landscape}
\begin{table*}[h!]
\centering
\caption{The platform designs in TikTok and Taobao's livestream shopping that mediated malicious selling strategies.}
\label{tab:designs}
\resizebox{1.4\textwidth}{!}{%
\begin{tabular}{p{0.18\linewidth}p{0.45\linewidth}|p{0.15\linewidth}|p{0.15\linewidth}|p{0.15\linewidth}|p{0.15\linewidth}}
\toprule
\textbf{Design} & \textbf{Explanation} & \textbf{Restrictive} & \textbf{Deceptive}& \textbf{Covert} & \textbf{Asymmetric}\\\midrule
Subscription & Viewers can click the ``subscribe'' button to follow the streamer and  receive future livestream session notifications. &  Forced Subscription & & &\\\midrule
Reminder tag & The streamer can set up a message tag to display important reminders (e.g., a subscribing reminder tag) or positive comments from viewers. The reminder message tag is enlarged and displayed on the screen (section 5 in~\autoref{fig:interface}\subref{fig:douyin}). &  Forced Subscription & & &\\\midrule
Sharing & Viewers can click the ``share'' button and share the livestream channel on social media or with other viewers (section 4 in~\autoref{fig:interface}\subref{fig:douyin}). &  Forced Endorsement & & &\\\midrule
Liking & Viewers can click the ``like'' button to express their like, enjoyment, or support to the streamer (section 4 in~\autoref{fig:interface}\subref{fig:douyin}). & Forced Endorsement & & &\\\midrule
Gifting & Viewers can click the ``gift'' button to send gifts to the streamer (section 4 in~\autoref{fig:interface}\subref{fig:douyin}). & Forced Endorsement & & &\\\midrule
Backend filtering tool & Streamers can use a backend filtering tool to select a group of viewers by a criteria (e.g., the viewers who have subscribed to the streamer). & Forced Subscription, Forced Endorsement & &  &\\\midrule
Notification message & The various events (e.g., purchasing, sold out, sending gifts, sending likes) are displayed on the screen above the chat channel (section 2 in~\autoref{fig:interface}\subref{fig:douyin}). &  & Fake Scarcity, Fake Social Proof & &\\\midrule
Product page & Viewers can see detailed information about products when they click on a product on the product page (section 6 in~\autoref{fig:interface}\subref{fig:shopcart}). & & Fake Scarcity, Fake Exclusive Pricing & &\\\midrule
Red pocket & The streamer can send an money to viewers in a red pocket. The red pocket will pop up on all viewers' screens. &  & Fake Exclusive Pricing & &\\\midrule
Coupon & The streamer can send coupons to viewers, which can be redeemed for a discount or a cash rebate. The coupon will pop up on all viewers' screens. & & Fake Exclusive Pricing & &\\\midrule
Countdown timer & The streamer can set up a countdown timer for a product sale. The countdown timer will be displayed on the screen. & & Fake Urgency & &\\\midrule
Back-ordered warning & A warning message periodically pops up to remind viewers to finalize the payment. & & Fake Urgency & &\\\midrule
Flash sale & The streamer can set up a time-limited sale. & & Fake Urgency & &\\\midrule
Beautify camera filter & The streamer can use a beautification camera filter to beautify their products and themselves. & & & Visual Misrepresentation &\\\midrule
Chat channel & Viewers can post text-based comments and see others comments in a publicly displayed section in real time (section 1 in~\autoref{fig:interface}\subref{fig:douyin}). & Forced Endorsement & Fake Scarcity, Fake Social Proof & & Retaining Customers, Egoistic Norms\\\bottomrule
\end{tabular}%
}
\end{table*}
\end{landscape}

\subsection{Viewer Challenges when Coping with Malicious Selling Strategies (RQ3)}
Participants reported multiple ways to cope with malicious selling strategies. The most frequent mechanism was to select professional streamers to watch due to their high reputation. A professional streamer is a salesperson or agent that acts as a liaison between product manufacturers and customers and takes a commission on the sale~\cite{StreamerProduct}. 
Participants noted that choosing brands and stores with a good reputation was a useful means to circumvent malicious selling strategies. In addition, participants suggested ``shopping around'' or comparing products in different livestream channels before buying. Participants also leveraged the chat channel to push back against malicious selling strategies. For example, they reported low-quality products, poor return service, or astroturfers in the chat to alert other viewers.

Participants reported four ways to deal with malicious selling strategies, however, not all participants considered these effective for some malicious strategies. For example, P5 planned to buy a pearl necklace for her mom. She watched different pearl livestream shopping channels and acquired the pearl knowledge introduced by the streamers. After about a week, she gained a sense of which necklaces were high-quality and found that some streamers had fake testimonials and did not tell the truth. It was also very time-consuming for her to acquire her knowledge. This suggests a need to understand the challenges inherent in defending against malicious selling strategies for viewers.

The interview-based data provided insights into the challenges of coping with malicious selling. These challenges were due to highly personalized stream procedures, malicious designs that impaired viewers, and the visual limitations of livestream for special products.

\subsubsection{\textbf{Highly Personalized Stream Procedures.}}
Stream procedures varied across platforms, streamers, and even products within one streamer's channel. There were no standards to guide streamers about how to sell products. Because the procedures were highly personalized and selected by streamers themselves, viewers encountered high costs when selecting products.

Within the same streamer's channel, non-standardized procedures including incomplete presentation, versatile scenarios aiming to target multi-type customers, and unpredictable stream orders led to problems. 
Unlike product pages on e-shopping websites, where product parameters and functions are presented in detail, product introductions on livestream were incomplete. The features or functions mentioned by streamers were highly selective and personalized. 
According to participants, some streamers prepared stream scripts in advance, which articulated which aspects of products should be emphasized in stream and which aspects should not be mentioned. 
Streamers also communicated with vendors and revised their stream scripts accordingly. P13 provided an example of an incomplete presentation. She said that in a livestream, it was easy to see the quality and appearance of a product but hard to see the expired date. The streamer only magnified the advantages of the products and did not tell customers the expired date. After receiving the product, customers would realize this.

When introducing products, streamers also obscured target users. This was related to the identified malicious strategy, Fuzzy Targeting. It was another type of incomplete presentation, i.e., streamers did not specify the target users which is usually a parameter on product pages on e-shopping websites. P7 compared the difference between streamers' product recommendations to customers and her recommendation to her friends. If a product was targeted at females aged 20 to 30, P7 would only consider whether the product was suitable for her friends. Alternately, streamers would convince customers that the product could also be used by their husbands and other family members.
\begin{quote}
    The streamer will say ``even if you don't want to use it, you can give it to your boyfriend or husband.'' The streamer has a very tricky logic loop, that is, she or he will make you feel that no matter who you are, you can always use this product... (P7)
\end{quote}
P13 mentioned another way of Fuzzy Targeting during livestream shopping. Instead of obscuring target users, streamers advocated that a product could be used in multiple scenarios. For example, P13 recalled that one streamer tried to convince customers that a coat was suitable for many contexts, such as home, school, and work. By imagining the contexts, customers would be convinced that it was good to buy this coat for different contexts.

Non-standardized stream procedures not only happened during single product introductions but also throughout the livestream's schedule. Several participants reported that stream schedules were not transparent or predictable. According to P7, livestream channels usually posted stream previews. However, they did not specify the sequence of streamed products or if the preview's sequences were different from the real one. They also applied strategies such as not forecasting the prices of the streamed products or forecasting only part of the streamed products. P7 explained that the purpose was to keep customers staying in the livestream channels for a longer time and thus buying more products. If customers quit livestream channels, there is no way to know when their desired products will be streamed. They need to wait in the channels, which requires a considerable time commitment.

P9 reported that after shopping in a livestream channel for three months, she could estimate when the product she wanted would be sold based on the forecast. But this only worked in that livestream channel because every product was usually displayed for five minutes in that channel. In other streamers' channels, the duration varied a lot. Participants also reported that many social media accounts were created to serve as livestream preview information hubs. These practises indicate the need for transparent livestream schedules for viewers.

Among different streamers channels, the non-standardized procedures influenced the effects of the products being display. Therefore, customers would like to choose streamers who followed a more standardized and professional way of selling products. P5 pointed out she used non-standardized selling procedures as a criterion to distinguish good streamers and bad streamers. Bad streamers were not good at setting up their livestream rooms or displaying products. There were no guidelines or tools to support them. She suggested that the livestream platforms should help streamers improve their service by providing ``intelligent modules,'' similar to standardized tools for helping vendors to set up product pages on e-shopping websites.  

Participants also reflected that there was no standardized procedure for streamers to follow. P5 reported that TikTok streamers had formed specific streaming styles but Taobao streamers did not. However, these TikTok styles were full of cyber slang and the purpose was mainly for entertaining customers. 

\subsubsection{\textbf{Malicious Designs Impairing Viewers.}}
Designs, particularly chat channels, impair viewers' capability as buyers due to a lack of identity and overwhelming numbers of comments.

The identity of viewers in livestreams was anonymous, which yielded astroturfers or fake users. Several participants reported that there were astroturfers, who were hired by streamers to post fake comments. The fake comments were usually showing the intention of buying or the action of ordering. When being asked whether there were fake comments and whether they could distinguish them from real ones, some participants told us there should be fake comments but they could not tell which ones were. P8 described his experience of identifying the astroturfers in a seal stone livestream channel accidentally:
\begin{quote}
    I wanted to buy a seal stone several days ago but someone raised the price a lot so I was not able to buy it. But two days later, I saw the same stone selling in the livestream channel again. I realized that the one who raised the price was an astroturfer. (P8)
\end{quote}

Astroturfers might also send negative comments to competitors' livestream channels. As far as we know, streamers can block users who have been identified as astroturfers. However, participants reflected that some streamers used fake comments and astroturfers as excuses to deny real customers' complaints. For example, P5 recalled that in one livestream channel, one viewer sent a negative comment. The streamer then blocked the viewer and claimed she was the competitor's astroturfer. At first, P5 believed the streamer's explanation. Later, another viewer sent a comment saying that \textit{``You blocked me? I switched to my friend's account just to prove I am not an astroturfer!''} P5 expressed her shock when she realized how the streamer could silence viewers.

In contrast to streamers, viewers have limited power to identify or block fake comments or astroturfers. 

Additionally, the chat channel design does not support conversations between customers. Thus, customers cannot share ideas within the livestream channel. P5 recalled an experience of exchanging products with another viewer in a livestream channel. Since the Taobao livestream platform does not allow direct private messaging between viewers, they needed to type in their Taobao account usernames in the chat channel and add each other as friends on Taobao. The chat channel does not allow copy and paste, which made communication much more difficult for P5. 
This cumbersome design seems to favor streamers. Participants expressed that they would rarely use the stream comments for asking questions. 

Participants (n = 3) also reported that comments often overwhelmed them. In some popular stream channels, there was usually an  influx of comments. Participants expressed that it was difficult to read these comments. As streamers often use the comment feature for promotional purposes, such as asking customers to send positive messages, e.g., ``I want it'' or ``I have ordered it'', streamers often orally highlighted positive comments during the livestream to induced viewers to buy products, but these comments were not helpful for viewers to understand products and added more complexity to the comment stream. 

\subsubsection{\textbf{The Visual Limitations of Livestream for Special Products.}}
Livestreams overcome some limitations of online shopping. because they create the feeling of presence. However, livestreaming still cannot fully replace the offline shopping experience for special products. Special products are usually non-industrialized and non-standardized products with large individual differences (e.g., jewellery and artwork). The malicious selling strategy Visual Mispresentation is related to this. To leverage the visual limitations of livestreams, streamers often use special light or embedded camera filters to beautify products. In contrast, during offline shopping, sellers manipulating products is more difficult.

P2 said that livestreams were not suitable for buying non-industrialized and non-standardized products. Because every product is unique, it is risky to buy this type of product through a livestream. He bought a pair of walnuts on livestream. The pair of walnuts was not food but an ``article for amusement.'' As it was a pair, it required the two walnuts to look almost the same. He said it was impossible to observe the product exactly as it would appear offline. This is the limitation of livestream shopping and the Visual Misrepresentation strategy amplifies the risky nature of buying non-industrialized products from livestream shopping:
\begin{quote}
It [the pair of walnuts] is not the kind of industrialized products so you are not able to see the product very intuitively only from streamers' livestream and pictures. You can only get a general idea, and that's based on the fact that everything he [the streamer] says is true. (P2)
\end{quote}

To add to this challenge, P8 also pointed out it was difficult to overcome the visual limitations of livestream shopping. P8 was interested in buying seal stones through livestream. He mentioned he spent a long time figuring out how to select high-quality seal stones. He usually spent a few days staying in a channel and observed the selling process to evaluate if it was worth buying. P8 shared that the light used by streamers could beautify the stones. Some streamers also hid flaws on stones and only showed the parts without flaws. However, without seeing the real product, he still failed to judge the quality of the products. The limitation brings a lot of time and money loss. 

In summary, the interview study revealed that viewers faced challenges in coping with streamers' malicious selling strategies and what factors led to these challenges. Next, we will discuss how to address the challenges.

\section{Discussion}
This research sought to provide an empirical understanding of malicious selling strategies in a newly emerging online shopping platform: livestream-based shopping. We sampled 40 livestream shopping videos as a data corpus and conducted qualitative coding analyses. We identified 16 selling malicious strategies. Out of the 16 strategies, nine were enhanced by the design of the livestream platforms themselves. We further summarized the 16 selling strategies into four categories (i.e., Restrictive, Deceptive, Covert, and Asymmetric). The interview study revealed viewers' awareness of malicious selling strategies, ways to cope with malicious strategies, and challenges when overcoming these malicious strategies. In the following section, we reflect on the main findings and discuss the regulation and cultural context of livestream commerce in China, the unbalanced power between viewers, streamers and platforms, as well as policy and design implications to counter malicious selling in livestream shopping. 

\subsection{A Reflection of Malicious Selling Strategies and Deceptive Designs in Livestream Shopping}

The recording coding process referred to existing typologies of deceptive designs. Although malicious selling strategies and deceptive designs are two different concepts, bad intentions or strategies behind selling strategies and deceptive designs shared some similarities. We borrowed codes such as ``urgency,'' ``scarcity,'' ``social proof,'' ``forced subscription,'' and ``nagging'' from the existing typologies~\cite{mathur2019dark,gray2018dark,crabtree1992doing}. More importantly, our study contributes new findings that are different from existing studies in two regards. 

First, we found that participants were more aware of Restrictive and Deceptive selling strategies than Covert and Asymmetric selling strategies. One reason could be that the frequency of these strategies were different. On a deeper level, the implementation of Covert and Asymmetric selling strategies relied more on sellers than platform design features or functions. Here, it seems that platform designs serve as checkpoints to verify if selling strategies were malicious or not. Without these checkpoints, malicious selling strategies might be more concealed. Thus, it is worth investigating if the absence of design features and functions will make the awareness of malicious selling strategies more difficult.

We also found that more than half of the malicious selling strategies were enhanced by platform designs. The chat channel supported more malicious selling strategies than other designs however, it was the only design used by viewers to fight against malicious selling. Restrictive selling strategies (e.g., Forced Endorsement, Forced Subscription) and Deceptive selling strategies (e.g., Fake Scarcity, Fake Social Proof, Fake Exclusive Pricing, and Fake Urgency) were more likely to be amplified by designs than others. Moreover, the Forced Endorsement and Forced Subscription strategies were amplified by multiple engagement designs such as subscriptions, reminder tags, sharing, liking, and gifting. These features were initially designed to engage viewers, however, they were maliciously used by streamers.

Second, compared to prior studies about malicious selling strategies during TV shopping and social shopping, our study identified new malicious strategies: Forced Subscription, Forced Endorsement, Playacting, Fuzzy Targeting, Sophistry, Retaining Customers, Egoistic Norms, and Disgracing Others. These strategies were enabled by the frequent and various interactions between a streamer and their viewers, such as the streamer's verbal communication and viewers' comments in the chat channel, sending gifts, and likes. This demonstrates how more attention needs to be paid to malicious selling strategies when the interaction between the seller and customers is more frequent and diverse.

One might argue that some malicious selling strategies identified in this work seem like traditional marketing strategies. In this work, strategies such as Retaining Customers and Playacting were similar to the content-based approach in Wongkitrungrueng's work~\cite{wongkitrungrueng2020live} and were not malicious. However, this work emphasizes that these livestream selling tactics were supported by platform functions or features and thus could manipulate viewers' decision-making more easily than offline marketing tactics. As Narayanan et al. argued, deceptive designs could be more aggressive than offline marketing strategies~\cite{narayanan2020dark}. Because designers can apply A/B tests to pick out the most manipulative design and make great profits for companies. As a result, ``even seemingly trivial changes to design elements can result in substantial differences in behavior''~\cite{narayanan2020dark}. It is thus important to evaluate the malicious side of selling strategies happening on livestream platforms.

\subsection{The Regulation and Cultural Context of Livestream Commerce in China}

We studied two popular livestream shopping platforms in China, where the regulation and cultural context of China might impact malicious selling and platform designs. To provide a better context for readers, we discuss how regulation and cultural context in China impacts livestream commerce, especially malicious selling strategies and deceptive designs.

First, from the regulation perspective, livestream e-commerce is regulated by state government departments in China. As previously mentioned, with the ambition of developing ``digital China,'' the state government has provided support to the internet economy, including livestream e-commerce. The government also encourages small businesses to leverage livestream to expand their business. Thus, many small business owners become streamers and sell their products via livestream. This leads to a variety of streamers on livestream platforms, ranging from professional streamers to amateur streamers from different industries (e.g., students, farmers). Thus, there are versatile and various selling strategies, including malicious selling strategies.

The state government tries to seek a balance between economic development and regulation. Currently, the regulations for livestream e-commerce in China are government policies rather than laws~\cite{zheng2021legal}. In 2020, the Cyberspace Administration of China announced that livestream platforms have to build a credit assessment list for livestreamers, where any livestreamer on the ``blacklist'' will be prohibited from using all livestreaming platforms~\cite{reuters2020livestreaming}. In 2021, the State Administration of Market Regulation said in a statement that all livestream platforms should ``quickly conduct self-control and comprehensive inspections'' on product quality, and should punish streamers who sell poor quality products or misleading advertising. These regulations demonstrate that livestream platforms take the main responsibility to monitor and regulate streamers' illegal or inappropriate behaviors. But there are no specific policies or laws to govern deceptive designs on livestream platforms. In the U.S., the DETOUR Act states that it is unlawful ``to design, modify, or manipulate a user interface with the purpose or substantial effect of obscuring, subverting, or impairing user autonomy, decision-making, or choice to obtain consent or user data''~\cite{warner2019senators}. Future research could further study the legality of deceptive designs in different online scenarios.

Second, from the cultural perspective, users in China prefer to use one app for many different functions~\cite{liu2021learn}. For example, WeChat has combined social media, e-commerce, payments, and many other functions into a self-contained ecosystem~\cite{davis2021westernizing}. Such integrated platforms are rarely seen in the U.S. (e.g., Facebook or Apple). Similarly, with livestream shopping, platforms try to include many features and functions in their interfaces. However, in the U.S. and other countries, the interface design for livestream shopping is much simpler. For example, Amazon has only four main features in its livestream interface: the streaming video, a list of products being sold, a chat area, and a reaction button where viewers can send ``likes'' to the streamer~\cite{amazonlive2021}. Platforms and streamers benefit from a greater degree of stability and opportunity in China and also experience a greater precarity in policy and regulation~\cite{cunningham2019china}. 

In China, the ecosystem of livestream commerce is becoming mature. It mainly consists of three roles: product and service providers (e.g., manufacturers, supply chains), marketers and retailers (e.g., livestream commerce platforms, streamers, online stores), and consumers~\cite{kpmg2020}. It might be fruitful in the future to investigate the roles of different stakeholders in this ecosystem, and to understand how they influence platform designs and regulation and how they perceive and cope with streamers' malicious selling strategies.

\subsection{The Imbalanced Power Between Viewers, Streamers, and Platforms}

Our research identified many negative behaviors from the streamer perspective. Although some of the streamers' malicious selling strategies were highly enhanced by system designs, we also identified seven malicious strategies that were purely implemented by streamers. These problems revealed standardized product stream procedures have not yet been established in livestream e-commerce. On the surface level, participants expressed that streamers were skilled in introducing products and persuading customers to buy. When we asked participants if they thought the ways or the strategies streamers used were malicious, they were not sure. Participants expressed that it was hard to differentiate exaggeration and deception. They reported cases where customers were
controlled by streamers or designs. We found that most  malicious strategies were directly or indirectly related to this. Compared to other e-shopping channels, customers were in a much more inferior position to streamers while livestream shopping. Streamers had a high level of freedom to decide how they sold products in livestream channels.

Prior work has analyzed viewers' negative behaviors during livestreams and how livestreaming platforms allow viewers to report inappropriate content~\cite{wohn2019volunteer}. For example, viewers reported trolling and being talkative in livestream channels. Given the unbalanced power between viewers and streamers, the platforms need to provide more ways for customers to combat malicious selling, such as collective reporting.

In the interview study, the participants mentioned several types of professional streamers, i.e., professional streamers, celebrity streamers, and influencer streamers. Some participants watched livestream e-commerce hosted by celebrities. Participants also watched influencers' livestreams and shopped in their channels. Before starting livestream selling, these influencers shared self-generated content (short videos) on TikTok and drew a large base of fans. Participants' trust and preference towards these professional streamers was due to their professionalism in choosing high-quality products and introducing products, and their authenticity within or outside of platforms. To fight against malicious selling, more professional streamers are needed.

\subsection{Design Implications}
These findings suggest that viewers' behaviors are constrained by the current designs of livestream platforms. In the livestream shopping setting, nine malicious selling strategies were enhanced by design features in livestream shopping channels. For instance, there were multiple notification features (e.g., purchase notification,  viewing notification, sold-out warning notification) that allowed streamers to create fake scarcity or urgency. The feeling of scarcity led to many impulse buying behaviors. Thus, streamers used such features and capitalized on the behaviors they resulted in. 
The chat channel, which was the main way for viewers to interact with the streamer and other viewers, was also manipulated by streamers to create fake scarcity impressions. Previous research has showed how synchronous chat facilitated knowledge sharing among viewers~\cite{wu2018danmaku}. However, as the streamer asked their viewers to post required content, such as their desired products in the chat channel, a viewer might only see thousands of ``wanting'' messages in the channel instead of product related knowledge shared by other viewers. Additionally, the chat channel did not support viewer-centric functions such as search, filter, and reply. The missing functions further constrained viewers' power to defend malicious selling behaviors. The designs of livestream platforms had a negative side when they were maliciously used by streamers. Gray and colleagues suggested that the intentions and motivations behind designs should be examined even if the designs were not deceptive designs~\cite{gray2020kind}. As our research suggests, engagement and communication designs should be studied with human behaviors (both streamers and viewers). An isolated investigation might overlook the potential negative side of the designs. Our study only investigated the features and functions level, however, in the future, an interesting direction could be to explore what properties of these features and functions can amplify malicious selling strategies.

Based on these findings, various design implications can be proposed to protect users from these malicious selling strategies and designs. First, it might be helpful to increase the transparency and viewers' autonomy in designs. For instance, the backend filtering tool was the key function for Restrictive selling strategies. Viewers need to know what filter mechanisms are available and used if they comply with such forced actions.
The chat channel also needs to be improved to enhance viewers' autonomy. Additional functions can be added to the chat channel, such as filter, search, or reply. 
Viewers should also be able to switch off message features (e.g., purchase message, viewing message, sold out message) on the screen so that they can focus on the product presentation and avoid being affected by others decisions. 
Second, it might be helpful to implement crowdsourced features to validate the information provided in livestream channels. For example, to counter the Fake Exclusive Pricing selling strategy, viewers can check prices across different e-commerce platforms or check other viewers' ratings or reviews to verify whether it was true or a false claimed exclusive pricing. 
We acknowledge such designs are unlikely to be adopted by the platform itself as it may be against its own business interests, however, third party developers may implement such designs as system add-ons to support viewers.

One general finding is that streamer's selling behaviors were loosely regulated. Even if they had bragged or promised something during the real-time streaming session, it was very hard for buyers to double-check whether or not it was true in the moment, and it was even more difficult to go back to what the streamer said after the session ended.
For example, a streamer may promise an unconditional return policy during the livestream session, but the actual return policy written on the product information page was conditional. Since malicious selling strategies relied more on streamers' sale skills and styles rather than platform features and functions, the platform and regulators should make and enforce policies to regulate streamers' selling practices similar to how they would regulate an offline salesperson. For example, it is important to determine if forced actions are legal or necessary for customers to purchase products during livestream shopping. Platforms should also implement machine learning-based solutions to automatically transcribe and record what the streamer is saying during the livestream session to extract keywords  from the transcripts and cross-reference the product information page (e.g., return policy, brand, and manufacture info, etc.). If there is a mismatch, the system may pop up a real-time alert to the viewers. Viewers can also use recorded transcripts as evidence in post-sale customer service.

\subsection{Limitations and Future Work}
Our study only examined two livestream shopping platforms---Taobao and TikTok and the interview participants were mainly users of these two platforms. The findings of malicious selling strategies may thus not be generalizable to other platforms given their different platform policies and designs. 

The current work is also descriptive. The qualitative analysis and the interview study only identified the presence of malicious selling strategies. Future work could use the taxonomy developed in this study to investigate the prevalence of various malicious selling strategies or run experiments to evaluate if malicious selling strategies could lead to financial loss of customers.

Additionally, we only recorded 40 livestream sessions over a short period. In the future, we plan to conduct a longitudinal analysis, that is, observing several livestream shopping channels as a real viewer over a long term. We will also conduct an interview study for streamers to understand their motivations and attitudes.

\section{Conclusion}
This research sought to understand the malicious selling strategies that are used on livestream shopping platforms and how platform designs support these malicious behaviors. These findings contribute to the broader HCI discussion about how to counter malicious strategies and deceptive designs, and to protect target users' welfare while platforms have the motivation and power to adopt more and more deceptive designs. We welcome researchers from various backgrounds to join our effort to push forward this prominent research agenda.




\bibliographystyle{ACM-Reference-Format}
\bibliography{0-reference,sample-base}


\begin{thebibliography}{88}


\ifx \showCODEN    \undefined \def \showCODEN     #1{\unskip}     \fi
\ifx \showDOI      \undefined \def \showDOI       #1{#1}\fi
\ifx \showISBNx    \undefined \def \showISBNx     #1{\unskip}     \fi
\ifx \showISBNxiii \undefined \def \showISBNxiii  #1{\unskip}     \fi
\ifx \showISSN     \undefined \def \showISSN      #1{\unskip}     \fi
\ifx \showLCCN     \undefined \def \showLCCN      #1{\unskip}     \fi
\ifx \shownote     \undefined \def \shownote      #1{#1}          \fi
\ifx \showarticletitle \undefined \def \showarticletitle #1{#1}   \fi
\ifx \showURL      \undefined \def \showURL       {\relax}        \fi
\providecommand\bibfield[2]{#2}
\providecommand\bibinfo[2]{#2}
\providecommand\natexlab[1]{#1}
\providecommand\showeprint[2][]{arXiv:#2}

\bibitem[\protect\citeauthoryear{Aggarwal and Vaidyanathan}{Aggarwal and
  Vaidyanathan}{2003}]%
        {aggarwal2003use}
\bibfield{author}{\bibinfo{person}{Praveen Aggarwal} {and}
  \bibinfo{person}{Rajiv Vaidyanathan}.} \bibinfo{year}{2003}\natexlab{}.
\newblock \showarticletitle{Use it or lose it: purchase acceleration effects of
  time-limited promotions}.
\newblock \bibinfo{journal}{\emph{Journal of Consumer Behaviour: An
  International Research Review}} \bibinfo{volume}{2}, \bibinfo{number}{4}
  (\bibinfo{year}{2003}), \bibinfo{pages}{393--403}.
\newblock


\bibitem[\protect\citeauthoryear{{Amazon Live}}{{Amazon Live}}{2021}]%
        {amazonlive2021}
\bibfield{author}{\bibinfo{person}{{Amazon Live}}.}
  \bibinfo{year}{2021}\natexlab{}.
\newblock \bibinfo{title}{Amazon Live}.
\newblock
\newblock
\urldef\tempurl%
\url{https://www.amazon.com/live}
\showURL{%
Retrieved 7/8/2021 from \tempurl}


\bibitem[\protect\citeauthoryear{Auter and Moore}{Auter and Moore}{1993}]%
        {auter1993buying}
\bibfield{author}{\bibinfo{person}{Philip~J Auter} {and} \bibinfo{person}{Roy~L
  Moore}.} \bibinfo{year}{1993}\natexlab{}.
\newblock \showarticletitle{Buying from a friend: A content analysis of two
  teleshopping programs}.
\newblock \bibinfo{journal}{\emph{Journalism Quarterly}} \bibinfo{volume}{70},
  \bibinfo{number}{2} (\bibinfo{year}{1993}), \bibinfo{pages}{425--436}.
\newblock


\bibitem[\protect\citeauthoryear{Benedicktus, Brady, Darke, and
  Voorhees}{Benedicktus et~al\mbox{.}}{2010}]%
        {benedicktus2010conveying}
\bibfield{author}{\bibinfo{person}{Ray~L Benedicktus},
  \bibinfo{person}{Michael~K Brady}, \bibinfo{person}{Peter~R Darke}, {and}
  \bibinfo{person}{Clay~M Voorhees}.} \bibinfo{year}{2010}\natexlab{}.
\newblock \showarticletitle{Conveying trustworthiness to online consumers:
  Reactions to consensus, physical store presence, brand familiarity, and
  generalized suspicion}.
\newblock \bibinfo{journal}{\emph{Journal of Retailing}} \bibinfo{volume}{86},
  \bibinfo{number}{4} (\bibinfo{year}{2010}), \bibinfo{pages}{322--335}.
\newblock


\bibitem[\protect\citeauthoryear{Brignull}{Brignull}{2018}]%
        {brignull2018dark}
\bibfield{author}{\bibinfo{person}{Harry Brignull}.}
  \bibinfo{year}{2018}\natexlab{}.
\newblock \bibinfo{title}{Types of deceptive design}.
\newblock
\newblock
\urldef\tempurl%
\url{https://www.deceptive.design/types}
\showURL{%
Retrieved 05/26/2022 from \tempurl}


\bibitem[\protect\citeauthoryear{Brignull, Miquel, Rosenberg, and
  Offer}{Brignull et~al\mbox{.}}{2015}]%
        {brignull2015dark}
\bibfield{author}{\bibinfo{person}{Harry Brignull}, \bibinfo{person}{Marc
  Miquel}, \bibinfo{person}{Jeremy Rosenberg}, {and} \bibinfo{person}{James
  Offer}.} \bibinfo{year}{2015}\natexlab{}.
\newblock \bibinfo{title}{Dark Patterns-User Interfaces Designed to Trick
  People}.
\newblock
\newblock
\urldef\tempurl%
\url{https://www.theverge.com/2013/8/29/4640308/dark-patterns-inside-the-interfaces-designed-to-trick-you}
\showURL{%
Retrieved 7/13/2021 from \tempurl}


\bibitem[\protect\citeauthoryear{Burgess}{Burgess}{2003}]%
        {burgess2003comparison}
\bibfield{author}{\bibinfo{person}{Brigitte Burgess}.}
  \bibinfo{year}{2003}\natexlab{}.
\newblock \showarticletitle{A comparison of TV home shoppers based on risk
  perception}.
\newblock \bibinfo{journal}{\emph{Journal of Fashion Marketing and Management:
  An International Journal}} (\bibinfo{year}{2003}).
\newblock


\bibitem[\protect\citeauthoryear{Burgess and Drake}{Burgess and Drake}{1995}]%
        {burgess1995television}
\bibfield{author}{\bibinfo{person}{B Burgess} {and} \bibinfo{person}{MF
  Drake}.} \bibinfo{year}{1995}\natexlab{}.
\newblock \showarticletitle{Television home shopping: A preliminary study}.
\newblock \bibinfo{journal}{\emph{Research in the Distributive Trades}}
  (\bibinfo{year}{1995}), \bibinfo{pages}{c2}.
\newblock


\bibitem[\protect\citeauthoryear{Cai and Wohn}{Cai and Wohn}{2019}]%
        {cai2019live}
\bibfield{author}{\bibinfo{person}{Jie Cai} {and}
  \bibinfo{person}{Donghee~Yvette Wohn}.} \bibinfo{year}{2019}\natexlab{}.
\newblock \showarticletitle{Live streaming commerce: Uses and gratifications
  approach to understanding consumers' motivations}. In
  \bibinfo{booktitle}{\emph{Proceedings of the 52nd Hawaii International
  Conference on System Sciences}}.
\newblock


\bibitem[\protect\citeauthoryear{Cai, Wohn, Mittal, and Sureshbabu}{Cai
  et~al\mbox{.}}{2018}]%
        {cai2018utilitarian}
\bibfield{author}{\bibinfo{person}{Jie Cai}, \bibinfo{person}{Donghee~Yvette
  Wohn}, \bibinfo{person}{Ankit Mittal}, {and} \bibinfo{person}{Dhanush
  Sureshbabu}.} \bibinfo{year}{2018}\natexlab{}.
\newblock \showarticletitle{Utilitarian and hedonic motivations for live
  streaming shopping}. In \bibinfo{booktitle}{\emph{Proceedings of the 2018 ACM
  international conference on interactive experiences for TV and online
  video}}. \bibinfo{pages}{81--88}.
\newblock


\bibitem[\protect\citeauthoryear{Cassidy}{Cassidy}{2020}]%
        {walmart2020tiktok}
\bibfield{author}{\bibinfo{person}{Tabitha Cassidy}.}
  \bibinfo{year}{2020}\natexlab{}.
\newblock \bibinfo{title}{Walmart pilots in-app, live-stream shopping on
  TikTok}.
\newblock
\newblock
\urldef\tempurl%
\url{https://www.digitalcommerce360.com/2020/12/18/walmart-pilots-in-app-live-stream-shopping-on-tiktok/}
\showURL{%
Retrieved 12/26/2020 from \tempurl}


\bibitem[\protect\citeauthoryear{Chen, Hu, Lu, and Hong}{Chen
  et~al\mbox{.}}{2019b}]%
        {chen2019everyone}
\bibfield{author}{\bibinfo{person}{Cheng Chen}, \bibinfo{person}{Yuheng Hu},
  \bibinfo{person}{Yingda Lu}, {and} \bibinfo{person}{Yili Hong}.}
  \bibinfo{year}{2019}\natexlab{b}.
\newblock \showarticletitle{Everyone can be a star: Quantifying grassroots
  online sellers' live streaming effects on product sales}.
\newblock  (\bibinfo{year}{2019}).
\newblock


\bibitem[\protect\citeauthoryear{Chen, Zhao, and Wang}{Chen
  et~al\mbox{.}}{2020b}]%
        {chen2020livestreaming}
\bibfield{author}{\bibinfo{person}{Chun-Der Chen}, \bibinfo{person}{Qun Zhao},
  {and} \bibinfo{person}{Jin-Long Wang}.} \bibinfo{year}{2020}\natexlab{b}.
\newblock \showarticletitle{How livestreaming increases product sales: role of
  trust transfer and elaboration likelihood model}.
\newblock \bibinfo{journal}{\emph{Behaviour \& Information Technology}}
  (\bibinfo{year}{2020}), \bibinfo{pages}{1--16}.
\newblock


\bibitem[\protect\citeauthoryear{Chen, Freeman, and Balakrishnan}{Chen
  et~al\mbox{.}}{2019a}]%
        {chen2019integrating}
\bibfield{author}{\bibinfo{person}{Di~(Laura) Chen}, \bibinfo{person}{Dustin
  Freeman}, {and} \bibinfo{person}{Ravin Balakrishnan}.}
  \bibinfo{year}{2019}\natexlab{a}.
\newblock \showarticletitle{Integrating Multimedia Tools to Enrich Interactions
  in Live Streaming for Language Learning}. In
  \bibinfo{booktitle}{\emph{Proceedings of the 2019 CHI Conference on Human
  Factors in Computing Systems}} (Glasgow, Scotland Uk)
  \emph{(\bibinfo{series}{CHI '19})}. \bibinfo{publisher}{Association for
  Computing Machinery}, \bibinfo{address}{New York, NY, USA},
  \bibinfo{pages}{1–14}.
\newblock
\showISBNx{9781450359702}
\urldef\tempurl%
\url{https://doi.org/10.1145/3290605.3300668}
\showDOI{\tempurl}


\bibitem[\protect\citeauthoryear{Chen, Iyer, and Padmanabhan}{Chen
  et~al\mbox{.}}{2002}]%
        {chen2002referral}
\bibfield{author}{\bibinfo{person}{Yuxin Chen}, \bibinfo{person}{Ganesh Iyer},
  {and} \bibinfo{person}{V Padmanabhan}.} \bibinfo{year}{2002}\natexlab{}.
\newblock \showarticletitle{Referral infomediaries}.
\newblock \bibinfo{journal}{\emph{Marketing Science}} \bibinfo{volume}{21},
  \bibinfo{number}{4} (\bibinfo{year}{2002}), \bibinfo{pages}{412--434}.
\newblock


\bibitem[\protect\citeauthoryear{Chen, Cao, Xu, Cheng, Wang, and Li}{Chen
  et~al\mbox{.}}{2020a}]%
        {chen2020understanding}
\bibfield{author}{\bibinfo{person}{Zhilong Chen}, \bibinfo{person}{Hancheng
  Cao}, \bibinfo{person}{Fengli Xu}, \bibinfo{person}{Mengjie Cheng},
  \bibinfo{person}{Tao Wang}, {and} \bibinfo{person}{Yong Li}.}
  \bibinfo{year}{2020}\natexlab{a}.
\newblock \showarticletitle{Understanding the Role of Intermediaries in Online
  Social E-commerce: An Exploratory Study of Beidian}.
\newblock \bibinfo{journal}{\emph{Proceedings of the ACM on Human-Computer
  Interaction}} \bibinfo{volume}{4}, \bibinfo{number}{CSCW2}
  (\bibinfo{year}{2020}), \bibinfo{pages}{1--24}.
\newblock


\bibitem[\protect\citeauthoryear{Cheng and Yang}{Cheng and Yang}{2020}]%
        {traction2020}
\bibfield{author}{\bibinfo{person}{Xian Cheng} {and} \bibinfo{person}{Jin
  Yang}.} \bibinfo{year}{2020}\natexlab{}.
\newblock \bibinfo{title}{Livestream e-commerce gains traction in China: Is it
  worth the hype?}
\newblock
\newblock
\urldef\tempurl%
\url{https://news.cgtn.com/news/2020-05-08/Livestream-e-commerce-gains-traction-in-China-Is-it-worth-the-hype--QjRRqc2kIo/index.html}
\showURL{%
Retrieved 1/14/2021 from \tempurl}


\bibitem[\protect\citeauthoryear{{China Marketing Insights}}{{China Marketing
  Insights}}{2020}]%
        {insights2020xinba}
\bibfield{author}{\bibinfo{person}{{China Marketing Insights}}.}
  \bibinfo{year}{2020}\natexlab{}.
\newblock \bibinfo{title}{Top Kuaishou Live Streamer Xinba Caught Selling Fake
  Products}.
\newblock
\newblock
\urldef\tempurl%
\url{http://chinamktginsights.com/top-kuaishou-live-streamer-xinba-caught-selling-fake-products/}
\showURL{%
Retrieved 1/14/2021 from \tempurl}


\bibitem[\protect\citeauthoryear{Cialdini}{Cialdini}{2009}]%
        {cialdini2009influence}
\bibfield{author}{\bibinfo{person}{Robert~B Cialdini}.}
  \bibinfo{year}{2009}\natexlab{}.
\newblock \bibinfo{booktitle}{\emph{Influence: Science and practice}}.
  Vol.~\bibinfo{volume}{4}.
\newblock \bibinfo{publisher}{Pearson education Boston, MA}.
\newblock


\bibitem[\protect\citeauthoryear{Conti and Sobiesk}{Conti and Sobiesk}{2010}]%
        {conti2010malicious}
\bibfield{author}{\bibinfo{person}{Gregory Conti} {and} \bibinfo{person}{Edward
  Sobiesk}.} \bibinfo{year}{2010}\natexlab{}.
\newblock \showarticletitle{Malicious interface design: exploiting the user}.
  In \bibinfo{booktitle}{\emph{Proceedings of the 19th international conference
  on World wide web}}. \bibinfo{pages}{271--280}.
\newblock


\bibitem[\protect\citeauthoryear{Cook}{Cook}{2000}]%
        {cook2000consumer}
\bibfield{author}{\bibinfo{person}{Judi~Puritz Cook}.}
  \bibinfo{year}{2000}\natexlab{}.
\newblock \showarticletitle{Consumer culture and television home shopping
  programming: an examination of the sales discourse}.
\newblock \bibinfo{journal}{\emph{Mass Communication \& Society}}
  \bibinfo{volume}{3}, \bibinfo{number}{4} (\bibinfo{year}{2000}),
  \bibinfo{pages}{373--391}.
\newblock


\bibitem[\protect\citeauthoryear{Crabtree and Miller}{Crabtree and
  Miller}{1992}]%
        {crabtree1992doing}
\bibfield{author}{\bibinfo{person}{Benjamin~F Crabtree} {and}
  \bibinfo{person}{William~L Miller}.} \bibinfo{year}{1992}\natexlab{}.
\newblock \showarticletitle{Doing qualitative research.}. In
  \bibinfo{booktitle}{\emph{Annual North American Primary Care Research Group
  Meeting, 19th, May, 1989, Quebec, PQ, Canada}}. Sage Publications, Inc.
\newblock


\bibitem[\protect\citeauthoryear{Creswell and Poth}{Creswell and Poth}{2016}]%
        {creswell2016qualitative}
\bibfield{author}{\bibinfo{person}{John~W Creswell} {and}
  \bibinfo{person}{Cheryl~N Poth}.} \bibinfo{year}{2016}\natexlab{}.
\newblock \bibinfo{booktitle}{\emph{Qualitative inquiry and research design:
  Choosing among five approaches}}.
\newblock \bibinfo{publisher}{Sage publications}.
\newblock


\bibitem[\protect\citeauthoryear{Cunningham, Craig, and Lv}{Cunningham
  et~al\mbox{.}}{2019}]%
        {cunningham2019china}
\bibfield{author}{\bibinfo{person}{Stuart Cunningham}, \bibinfo{person}{David
  Craig}, {and} \bibinfo{person}{Junyi Lv}.} \bibinfo{year}{2019}\natexlab{}.
\newblock \showarticletitle{China's livestreaming industry: platforms,
  politics, and precarity}.
\newblock \bibinfo{journal}{\emph{International Journal of Cultural Studies}}
  \bibinfo{volume}{22}, \bibinfo{number}{6} (\bibinfo{year}{2019}),
  \bibinfo{pages}{719--736}.
\newblock


\bibitem[\protect\citeauthoryear{{Daily Curiosity}}{{Daily Curiosity}}{2017}]%
        {Shopwindow}
\bibfield{author}{\bibinfo{person}{{Daily Curiosity}}.}
  \bibinfo{year}{2017}\natexlab{}.
\newblock \bibinfo{booktitle}{\emph{TikTok Adds ``Shopwindow''}}.
\newblock \bibinfo{type}{{T}echnical {R}eport}. \bibinfo{institution}{Daily
  Curiosity}, \bibinfo{address}{China}.
\newblock
\urldef\tempurl%
\url{http://web.archive.org/web/20080207010024/http://www.808multimedia.com/winnt/kernel.htm}
\showURL{%
\tempurl}


\bibitem[\protect\citeauthoryear{Davis and Xiao}{Davis and Xiao}{2021}]%
        {davis2021westernizing}
\bibfield{author}{\bibinfo{person}{Mark Davis} {and} \bibinfo{person}{Jian
  Xiao}.} \bibinfo{year}{2021}\natexlab{}.
\newblock \showarticletitle{De-westernizing platform studies: History and
  logics of Chinese and US platforms}.
\newblock \bibinfo{journal}{\emph{International Journal of Communication}}
  \bibinfo{volume}{15} (\bibinfo{year}{2021}), \bibinfo{pages}{20}.
\newblock


\bibitem[\protect\citeauthoryear{Deterding, Stenros, and Montola}{Deterding
  et~al\mbox{.}}{2020}]%
        {deterding2020against}
\bibfield{author}{\bibinfo{person}{Christoph~Sebastian Deterding},
  \bibinfo{person}{Jaakko Stenros}, {and} \bibinfo{person}{Markus Montola}.}
  \bibinfo{year}{2020}\natexlab{}.
\newblock \showarticletitle{Against "Dark Game Design Patterns"}. In
  \bibinfo{booktitle}{\emph{DiGRA'20-Abstract Proceedings of the 2020 DiGRA
  International Conference}}. York.
\newblock


\bibitem[\protect\citeauthoryear{Di~Geronimo, Braz, Fregnan, Palomba, and
  Bacchelli}{Di~Geronimo et~al\mbox{.}}{2020}]%
        {di2020ui}
\bibfield{author}{\bibinfo{person}{Linda Di~Geronimo}, \bibinfo{person}{Larissa
  Braz}, \bibinfo{person}{Enrico Fregnan}, \bibinfo{person}{Fabio Palomba},
  {and} \bibinfo{person}{Alberto Bacchelli}.} \bibinfo{year}{2020}\natexlab{}.
\newblock \showarticletitle{UI dark patterns and where to find them: a study on
  mobile applications and user perception}. In
  \bibinfo{booktitle}{\emph{Proceedings of the 2020 CHI Conference on Human
  Factors in Computing Systems}}. \bibinfo{pages}{1--14}.
\newblock


\bibitem[\protect\citeauthoryear{Faas, Dombrowski, Young, and Miller}{Faas
  et~al\mbox{.}}{2018}]%
        {faas2018watch}
\bibfield{author}{\bibinfo{person}{Travis Faas}, \bibinfo{person}{Lynn
  Dombrowski}, \bibinfo{person}{Alyson Young}, {and} \bibinfo{person}{Andrew~D.
  Miller}.} \bibinfo{year}{2018}\natexlab{}.
\newblock \showarticletitle{Watch Me Code: Programming Mentorship Communities
  on Twitch.Tv}.
\newblock \bibinfo{journal}{\emph{Proc. ACM Hum.-Comput. Interact.}}
  \bibinfo{volume}{2}, \bibinfo{number}{CSCW}, Article \bibinfo{articleno}{50}
  (\bibinfo{date}{Nov.} \bibinfo{year}{2018}), \bibinfo{numpages}{18}~pages.
\newblock
\urldef\tempurl%
\url{https://doi.org/10.1145/3274319}
\showDOI{\tempurl}


\bibitem[\protect\citeauthoryear{{Financing China}}{{Financing China}}{2019}]%
        {StreamerProduct}
\bibfield{author}{\bibinfo{person}{{Financing China}}.}
  \bibinfo{year}{2019}\natexlab{}.
\newblock \bibinfo{title}{Business Model of Livestream Shopping}.
\newblock
\newblock
\urldef\tempurl%
\url{https://36kr.com/p/1724744613889}
\showURL{%
Retrieved 1/14/2021 from \tempurl}


\bibitem[\protect\citeauthoryear{Glickman, McKenzie, Seering, Moeller, and
  Hammer}{Glickman et~al\mbox{.}}{2018}]%
        {clickman2018design}
\bibfield{author}{\bibinfo{person}{Seth Glickman}, \bibinfo{person}{Nathan
  McKenzie}, \bibinfo{person}{Joseph Seering}, \bibinfo{person}{Rachel
  Moeller}, {and} \bibinfo{person}{Jessica Hammer}.}
  \bibinfo{year}{2018}\natexlab{}.
\newblock \showarticletitle{Design Challenges for Livestreamed Audience
  Participation Games}. In \bibinfo{booktitle}{\emph{Proceedings of the 2018
  Annual Symposium on Computer-Human Interaction in Play}} (Melbourne, VIC,
  Australia) \emph{(\bibinfo{series}{CHI PLAY '18})}.
  \bibinfo{publisher}{Association for Computing Machinery},
  \bibinfo{address}{New York, NY, USA}, \bibinfo{pages}{187–199}.
\newblock
\showISBNx{9781450356244}
\urldef\tempurl%
\url{https://doi.org/10.1145/3242671.3242708}
\showDOI{\tempurl}


\bibitem[\protect\citeauthoryear{Grant, Guthrie, and Ball-Rokeach}{Grant
  et~al\mbox{.}}{1991}]%
        {grant1991television}
\bibfield{author}{\bibinfo{person}{August~E Grant}, \bibinfo{person}{K~Kendall
  Guthrie}, {and} \bibinfo{person}{Sandra~J Ball-Rokeach}.}
  \bibinfo{year}{1991}\natexlab{}.
\newblock \showarticletitle{Television shopping: A media system dependency
  perspective}.
\newblock \bibinfo{journal}{\emph{Communication Research}}
  \bibinfo{volume}{18}, \bibinfo{number}{6} (\bibinfo{year}{1991}),
  \bibinfo{pages}{773--798}.
\newblock


\bibitem[\protect\citeauthoryear{Gray, Chivukula, and Lee}{Gray
  et~al\mbox{.}}{2020}]%
        {gray2020kind}
\bibfield{author}{\bibinfo{person}{Colin~M Gray}, \bibinfo{person}{Shruthi~Sai
  Chivukula}, {and} \bibinfo{person}{Ahreum Lee}.}
  \bibinfo{year}{2020}\natexlab{}.
\newblock \showarticletitle{What Kind of Work Do ``Asshole Designers'' Create?
  Describing Properties of Ethical Concern on Reddit}. In
  \bibinfo{booktitle}{\emph{Proceedings of the 2020 ACM Designing Interactive
  Systems Conference}}. \bibinfo{pages}{61--73}.
\newblock


\bibitem[\protect\citeauthoryear{Gray, Kou, Battles, Hoggatt, and Toombs}{Gray
  et~al\mbox{.}}{2018}]%
        {gray2018dark}
\bibfield{author}{\bibinfo{person}{Colin~M Gray}, \bibinfo{person}{Yubo Kou},
  \bibinfo{person}{Bryan Battles}, \bibinfo{person}{Joseph Hoggatt}, {and}
  \bibinfo{person}{Austin~L Toombs}.} \bibinfo{year}{2018}\natexlab{}.
\newblock \showarticletitle{The dark (patterns) side of UX design}. In
  \bibinfo{booktitle}{\emph{Proceedings of the 2018 CHI Conference on Human
  Factors in Computing Systems}}. \bibinfo{pages}{1--14}.
\newblock


\bibitem[\protect\citeauthoryear{Greenberg}{Greenberg}{2016}]%
        {greenberg2016interaction}
\bibfield{author}{\bibinfo{person}{Jack Greenberg}.}
  \bibinfo{year}{2016}\natexlab{}.
\newblock \showarticletitle{Interaction between audience and game players
  during live streaming of games}.
\newblock  (\bibinfo{year}{2016}).
\newblock


\bibitem[\protect\citeauthoryear{Hajli}{Hajli}{2015}]%
        {hajli2015social}
\bibfield{author}{\bibinfo{person}{Nick Hajli}.}
  \bibinfo{year}{2015}\natexlab{}.
\newblock \showarticletitle{Social commerce constructs and consumer's intention
  to buy}.
\newblock \bibinfo{journal}{\emph{International Journal of Information
  Management}} \bibinfo{volume}{35}, \bibinfo{number}{2}
  (\bibinfo{year}{2015}), \bibinfo{pages}{183--191}.
\newblock


\bibitem[\protect\citeauthoryear{Hamilton, Karahalios, Sandvig, and
  Eslami}{Hamilton et~al\mbox{.}}{2014b}]%
        {hamilton2014path}
\bibfield{author}{\bibinfo{person}{Kevin Hamilton}, \bibinfo{person}{Karrie
  Karahalios}, \bibinfo{person}{Christian Sandvig}, {and}
  \bibinfo{person}{Motahhare Eslami}.} \bibinfo{year}{2014}\natexlab{b}.
\newblock \showarticletitle{A Path to Understanding the Effects of Algorithm
  Awareness}. In \bibinfo{booktitle}{\emph{CHI '14 Extended Abstracts on Human
  Factors in Computing Systems}} (Toronto, Ontario, Canada)
  \emph{(\bibinfo{series}{CHI EA '14})}. \bibinfo{publisher}{Association for
  Computing Machinery}, \bibinfo{address}{New York, NY, USA},
  \bibinfo{pages}{631–642}.
\newblock
\showISBNx{9781450324748}
\urldef\tempurl%
\url{https://doi.org/10.1145/2559206.2578883}
\showDOI{\tempurl}


\bibitem[\protect\citeauthoryear{Hamilton, Garretson, and Kerne}{Hamilton
  et~al\mbox{.}}{2014a}]%
        {hamilton2014streaming}
\bibfield{author}{\bibinfo{person}{William~A. Hamilton},
  \bibinfo{person}{Oliver Garretson}, {and} \bibinfo{person}{Andruid Kerne}.}
  \bibinfo{year}{2014}\natexlab{a}.
\newblock \showarticletitle{Streaming on Twitch: Fostering Participatory
  Communities of Play within Live Mixed Media}. In
  \bibinfo{booktitle}{\emph{Proceedings of the SIGCHI Conference on Human
  Factors in Computing Systems}} (Toronto, Ontario, Canada)
  \emph{(\bibinfo{series}{CHI '14})}. \bibinfo{publisher}{Association for
  Computing Machinery}, \bibinfo{address}{New York, NY, USA},
  \bibinfo{pages}{1315–1324}.
\newblock
\showISBNx{9781450324731}
\urldef\tempurl%
\url{https://doi.org/10.1145/2556288.2557048}
\showDOI{\tempurl}


\bibitem[\protect\citeauthoryear{Hamilton, Lupfer, Botello, Tesch, Stacy,
  Merrill, Williford, Bentley, and Kerne}{Hamilton et~al\mbox{.}}{2018}]%
        {hamilton2018collaborative}
\bibfield{author}{\bibinfo{person}{William~A. Hamilton}, \bibinfo{person}{Nic
  Lupfer}, \bibinfo{person}{Nicolas Botello}, \bibinfo{person}{Tyler Tesch},
  \bibinfo{person}{Alex Stacy}, \bibinfo{person}{Jeremy Merrill},
  \bibinfo{person}{Blake Williford}, \bibinfo{person}{Frank~R. Bentley}, {and}
  \bibinfo{person}{Andruid Kerne}.} \bibinfo{year}{2018}\natexlab{}.
\newblock \showarticletitle{Collaborative Live Media Curation: Shared Context
  for Participation in Online Learning}. In
  \bibinfo{booktitle}{\emph{Proceedings of the 2018 CHI Conference on Human
  Factors in Computing Systems}} (Montreal QC, Canada)
  \emph{(\bibinfo{series}{CHI '18})}. \bibinfo{publisher}{Association for
  Computing Machinery}, \bibinfo{address}{New York, NY, USA},
  \bibinfo{pages}{1–14}.
\newblock
\showISBNx{9781450356206}
\urldef\tempurl%
\url{https://doi.org/10.1145/3173574.3174129}
\showDOI{\tempurl}


\bibitem[\protect\citeauthoryear{Kapner}{Kapner}{2020}]%
        {kapner2020qvc}
\bibfield{author}{\bibinfo{person}{Suzanne Kapner}.}
  \bibinfo{year}{2020}\natexlab{}.
\newblock \bibinfo{title}{Levi's, Hilfiger Push a New Kind of Online Shopping.
  It Looks a Lot Like QVC.}
\newblock
\newblock
\urldef\tempurl%
\url{https://www.wsj.com/articles/levis-hilfiger-push-a-new-kind-of-online-shopping-it-looks-a-lot-like-qvc-11602424800}
\showURL{%
Retrieved 1/15/2020 from \tempurl}


\bibitem[\protect\citeauthoryear{Kaytoue, Silva, Cerf, Meira, and
  Ra\"{\i}ssi}{Kaytoue et~al\mbox{.}}{2012}]%
        {kaytoue2012watch}
\bibfield{author}{\bibinfo{person}{Mehdi Kaytoue}, \bibinfo{person}{Arlei
  Silva}, \bibinfo{person}{Lo\"{\i}c Cerf}, \bibinfo{person}{Wagner Meira},
  {and} \bibinfo{person}{Chedy Ra\"{\i}ssi}.} \bibinfo{year}{2012}\natexlab{}.
\newblock \showarticletitle{Watch Me Playing, i Am a Professional: A First
  Study on Video Game Live Streaming}. In \bibinfo{booktitle}{\emph{Proceedings
  of the 21st International Conference on World Wide Web}} (Lyon, France)
  \emph{(\bibinfo{series}{WWW '12 Companion})}. \bibinfo{publisher}{Association
  for Computing Machinery}, \bibinfo{address}{New York, NY, USA},
  \bibinfo{pages}{1181–1188}.
\newblock
\showISBNx{9781450312301}
\urldef\tempurl%
\url{https://doi.org/10.1145/2187980.2188259}
\showDOI{\tempurl}


\bibitem[\protect\citeauthoryear{Kempe-Cook, Sher, and Su}{Kempe-Cook
  et~al\mbox{.}}{2019}]%
        {cook2019behind}
\bibfield{author}{\bibinfo{person}{Lucas Kempe-Cook}, \bibinfo{person}{Stephen
  Tsung-Han Sher}, {and} \bibinfo{person}{Norman~Makoto Su}.}
  \bibinfo{year}{2019}\natexlab{}.
\newblock \showarticletitle{Behind the Voices: The Practice and Challenges of
  Esports Casters}. In \bibinfo{booktitle}{\emph{Proceedings of the 2019 CHI
  Conference on Human Factors in Computing Systems}} (Glasgow, Scotland Uk)
  \emph{(\bibinfo{series}{CHI '19})}. \bibinfo{publisher}{Association for
  Computing Machinery}, \bibinfo{address}{New York, NY, USA},
  \bibinfo{pages}{1–12}.
\newblock
\showISBNx{9781450359702}
\urldef\tempurl%
\url{https://doi.org/10.1145/3290605.3300795}
\showDOI{\tempurl}


\bibitem[\protect\citeauthoryear{Kim and Park}{Kim and Park}{2013}]%
        {kim2013effects}
\bibfield{author}{\bibinfo{person}{Sanghyun Kim} {and} \bibinfo{person}{Hyunsun
  Park}.} \bibinfo{year}{2013}\natexlab{}.
\newblock \showarticletitle{Effects of various characteristics of social
  commerce (s-commerce) on consumers' trust and trust performance}.
\newblock \bibinfo{journal}{\emph{International Journal of Information
  Management}} \bibinfo{volume}{33}, \bibinfo{number}{2}
  (\bibinfo{year}{2013}), \bibinfo{pages}{318--332}.
\newblock


\bibitem[\protect\citeauthoryear{Kline}{Kline}{2005}]%
        {kline2005interactive}
\bibfield{author}{\bibinfo{person}{Susan~L Kline}.}
  \bibinfo{year}{2005}\natexlab{}.
\newblock \showarticletitle{Interactive media systems: Influence strategies in
  television home shopping}.
\newblock \bibinfo{journal}{\emph{Text \& Talk}} \bibinfo{volume}{25},
  \bibinfo{number}{2} (\bibinfo{year}{2005}), \bibinfo{pages}{201--231}.
\newblock


\bibitem[\protect\citeauthoryear{{KPMG} and {Ali Research Institute}}{{KPMG}
  and {Ali Research Institute}}{2020}]%
        {kpmg2020}
\bibfield{author}{\bibinfo{person}{{KPMG}} {and} \bibinfo{person}{{Ali Research
  Institute}}.} \bibinfo{year}{2020}\natexlab{}.
\newblock \bibinfo{booktitle}{\emph{Live Streaming E-commerce is Stepping into
  the Trillion Era}}.
\newblock \bibinfo{type}{{T}echnical {R}eport} CN-MARKETS20-0005c.
  \bibinfo{institution}{KPMG}, \bibinfo{address}{China}.
\newblock
\urldef\tempurl%
\url{https://assets.kpmg/content/dam/kpmg/cn/pdf/zh/2020/10/live-streaming-e-commerce-towards-trillion-market.pdf}
\showURL{%
\tempurl}


\bibitem[\protect\citeauthoryear{Krippendorff}{Krippendorff}{2004}]%
        {krippendorff2004reliability}
\bibfield{author}{\bibinfo{person}{Klaus Krippendorff}.}
  \bibinfo{year}{2004}\natexlab{}.
\newblock \showarticletitle{Reliability in content analysis}.
\newblock \bibinfo{journal}{\emph{Human communication research}}
  \bibinfo{volume}{30}, \bibinfo{number}{3} (\bibinfo{year}{2004}),
  \bibinfo{pages}{411--433}.
\newblock


\bibitem[\protect\citeauthoryear{Lacey and Caudwell}{Lacey and
  Caudwell}{2019}]%
        {lacey2019cuteness}
\bibfield{author}{\bibinfo{person}{Cherie Lacey} {and}
  \bibinfo{person}{Catherine Caudwell}.} \bibinfo{year}{2019}\natexlab{}.
\newblock \showarticletitle{Cuteness as a `Dark Pattern' in Home Robots}. In
  \bibinfo{booktitle}{\emph{2019 14th ACM/IEEE International Conference on
  Human-Robot Interaction (HRI)}}. IEEE, \bibinfo{pages}{374--381}.
\newblock


\bibitem[\protect\citeauthoryear{Lessel, Vielhauer, and Kr\"{u}ger}{Lessel
  et~al\mbox{.}}{2017}]%
        {lessel2017expanding}
\bibfield{author}{\bibinfo{person}{Pascal Lessel}, \bibinfo{person}{Alexander
  Vielhauer}, {and} \bibinfo{person}{Antonio Kr\"{u}ger}.}
  \bibinfo{year}{2017}\natexlab{}.
\newblock \bibinfo{booktitle}{\emph{Expanding Video Game Live-Streams with
  Enhanced Communication Channels: A Case Study}}.
\newblock \bibinfo{publisher}{Association for Computing Machinery},
  \bibinfo{address}{New York, NY, USA}, \bibinfo{pages}{1571–1576}.
\newblock
\showISBNx{9781450346559}
\urldef\tempurl%
\url{https://doi.org/10.1145/3025453.3025708}
\showURL{%
\tempurl}


\bibitem[\protect\citeauthoryear{Li, Min, Hu, and Liu}{Li
  et~al\mbox{.}}{2020}]%
        {li2020understanding}
\bibfield{author}{\bibinfo{person}{Mengyi Li}, \bibinfo{person}{Qingfei Min},
  \bibinfo{person}{Lixia Hu}, {and} \bibinfo{person}{Zhiyong Liu}.}
  \bibinfo{year}{2020}\natexlab{}.
\newblock \showarticletitle{Understanding Live Streaming Shopping Intentions: A
  Vicarious Learning Perspective.}. In \bibinfo{booktitle}{\emph{PACIS}}.
  \bibinfo{pages}{108}.
\newblock


\bibitem[\protect\citeauthoryear{Liu}{Liu}{2021}]%
        {liu2021learn}
\bibfield{author}{\bibinfo{person}{Feifei Liu}.}
  \bibinfo{year}{2021}\natexlab{}.
\newblock \bibinfo{title}{Livestream Ecommerce: What We Can Learn from China}.
\newblock
\newblock
\urldef\tempurl%
\url{https://www.nngroup.com/articles/livestream-ecommerce-china/}
\showURL{%
Retrieved 7/2/2021 from \tempurl}


\bibitem[\protect\citeauthoryear{Lottridge, Bentley, Wheeler, Lee, Cheung, Ong,
  and Rowley}{Lottridge et~al\mbox{.}}{2017}]%
        {lottridge2017third}
\bibfield{author}{\bibinfo{person}{Danielle Lottridge}, \bibinfo{person}{Frank
  Bentley}, \bibinfo{person}{Matt Wheeler}, \bibinfo{person}{Jason Lee},
  \bibinfo{person}{Janet Cheung}, \bibinfo{person}{Katherine Ong}, {and}
  \bibinfo{person}{Cristy Rowley}.} \bibinfo{year}{2017}\natexlab{}.
\newblock \showarticletitle{Third-Wave Livestreaming: Teens' Long Form Selfie}.
  In \bibinfo{booktitle}{\emph{Proceedings of the 19th International Conference
  on Human-Computer Interaction with Mobile Devices and Services}} (Vienna,
  Austria) \emph{(\bibinfo{series}{MobileHCI '17})}.
  \bibinfo{publisher}{Association for Computing Machinery},
  \bibinfo{address}{New York, NY, USA}, Article \bibinfo{articleno}{20},
  \bibinfo{numpages}{12}~pages.
\newblock
\showISBNx{9781450350754}
\urldef\tempurl%
\url{https://doi.org/10.1145/3098279.3098540}
\showDOI{\tempurl}


\bibitem[\protect\citeauthoryear{Lu, Annett, Fan, and Wigdor}{Lu
  et~al\mbox{.}}{2019}]%
        {lu2019feel}
\bibfield{author}{\bibinfo{person}{Zhicong Lu}, \bibinfo{person}{Michelle
  Annett}, \bibinfo{person}{Mingming Fan}, {and} \bibinfo{person}{Daniel
  Wigdor}.} \bibinfo{year}{2019}\natexlab{}.
\newblock \showarticletitle{``I Feel It is My Responsibility to Stream'':
  Streaming and Engaging with Intangible Cultural Heritage through
  Livestreaming}. In \bibinfo{booktitle}{\emph{Proceedings of the 2019 CHI
  Conference on Human Factors in Computing Systems}} (Glasgow, Scotland Uk)
  \emph{(\bibinfo{series}{CHI '19})}. \bibinfo{publisher}{Association for
  Computing Machinery}, \bibinfo{address}{New York, NY, USA},
  \bibinfo{pages}{1–14}.
\newblock
\showISBNx{9781450359702}
\urldef\tempurl%
\url{https://doi.org/10.1145/3290605.3300459}
\showDOI{\tempurl}


\bibitem[\protect\citeauthoryear{Lu, Heo, and Wigdor}{Lu
  et~al\mbox{.}}{2018a}]%
        {lu2018streamwiki}
\bibfield{author}{\bibinfo{person}{Zhicong Lu}, \bibinfo{person}{Seongkook
  Heo}, {and} \bibinfo{person}{Daniel~J. Wigdor}.}
  \bibinfo{year}{2018}\natexlab{a}.
\newblock \showarticletitle{StreamWiki: Enabling Viewers of Knowledge Sharing
  Live Streams to Collaboratively Generate Archival Documentation for Effective
  In-Stream and Post Hoc Learning}.
\newblock \bibinfo{journal}{\emph{Proc. ACM Hum.-Comput. Interact.}}
  \bibinfo{volume}{2}, \bibinfo{number}{CSCW}, Article \bibinfo{articleno}{112}
  (\bibinfo{date}{Nov.} \bibinfo{year}{2018}), \bibinfo{numpages}{26}~pages.
\newblock
\urldef\tempurl%
\url{https://doi.org/10.1145/3274381}
\showDOI{\tempurl}


\bibitem[\protect\citeauthoryear{Lu, Xia, Heo, and Wigdor}{Lu
  et~al\mbox{.}}{2018b}]%
        {lu2018you}
\bibfield{author}{\bibinfo{person}{Zhicong Lu}, \bibinfo{person}{Haijun Xia},
  \bibinfo{person}{Seongkook Heo}, {and} \bibinfo{person}{Daniel Wigdor}.}
  \bibinfo{year}{2018}\natexlab{b}.
\newblock \showarticletitle{You Watch, You Give, and You Engage: A Study of
  Live Streaming Practices in China}. In \bibinfo{booktitle}{\emph{Proceedings
  of the 2018 CHI Conference on Human Factors in Computing Systems}} (Montreal
  QC, Canada) \emph{(\bibinfo{series}{CHI '18})}.
  \bibinfo{publisher}{Association for Computing Machinery},
  \bibinfo{address}{New York, NY, USA}, \bibinfo{pages}{1–13}.
\newblock
\showISBNx{9781450356206}
\urldef\tempurl%
\url{https://doi.org/10.1145/3173574.3174040}
\showDOI{\tempurl}


\bibitem[\protect\citeauthoryear{Luguri and Strahilevitz}{Luguri and
  Strahilevitz}{2019}]%
        {luguri2019shining}
\bibfield{author}{\bibinfo{person}{Jamie Luguri} {and} \bibinfo{person}{Lior
  Strahilevitz}.} \bibinfo{year}{2019}\natexlab{}.
\newblock \showarticletitle{Shining a light on dark patterns}.
\newblock \bibinfo{journal}{\emph{U of Chicago, Public Law Working Paper}}
  \bibinfo{number}{719} (\bibinfo{year}{2019}).
\newblock


\bibitem[\protect\citeauthoryear{Madhok and {CNN's Beijing bureau}}{Madhok and
  {CNN's Beijing bureau}}{2021}]%
        {cnn2021viya}
\bibfield{author}{\bibinfo{person}{Diksha Madhok} {and} \bibinfo{person}{{CNN's
  Beijing bureau}}.} \bibinfo{year}{2021}\natexlab{}.
\newblock \bibinfo{title}{China fines `live-streaming queen' Viya \$210 million
  for tax evasion}.
\newblock
\newblock
\urldef\tempurl%
\url{https://www.cnn.com/2021/12/21/tech/china-livestreamer-viya-fine-tax-evasion/index.html}
\showURL{%
Retrieved 6/16/2022 from \tempurl}


\bibitem[\protect\citeauthoryear{Maier and Harr}{Maier and Harr}{2020}]%
        {maier2020dark}
\bibfield{author}{\bibinfo{person}{Maximilian Maier} {and}
  \bibinfo{person}{Rikard Harr}.} \bibinfo{year}{2020}\natexlab{}.
\newblock \showarticletitle{Dark Design Patterns: An End-User Perspective}.
\newblock \bibinfo{journal}{\emph{Human Technology}} \bibinfo{volume}{16},
  \bibinfo{number}{2} (\bibinfo{year}{2020}).
\newblock


\bibitem[\protect\citeauthoryear{Mathur, Acar, Friedman, Lucherini, Mayer,
  Chetty, and Narayanan}{Mathur et~al\mbox{.}}{2019}]%
        {mathur2019dark}
\bibfield{author}{\bibinfo{person}{Arunesh Mathur}, \bibinfo{person}{Gunes
  Acar}, \bibinfo{person}{Michael~J Friedman}, \bibinfo{person}{Elena
  Lucherini}, \bibinfo{person}{Jonathan Mayer}, \bibinfo{person}{Marshini
  Chetty}, {and} \bibinfo{person}{Arvind Narayanan}.}
  \bibinfo{year}{2019}\natexlab{}.
\newblock \showarticletitle{Dark patterns at scale: Findings from a crawl of
  11K shopping websites}.
\newblock \bibinfo{journal}{\emph{Proceedings of the ACM on Human-Computer
  Interaction}} \bibinfo{volume}{3}, \bibinfo{number}{CSCW}
  (\bibinfo{year}{2019}), \bibinfo{pages}{1--32}.
\newblock


\bibitem[\protect\citeauthoryear{Mathur, Kshirsagar, and Mayer}{Mathur
  et~al\mbox{.}}{2021}]%
        {mathur2021makes}
\bibfield{author}{\bibinfo{person}{Arunesh Mathur}, \bibinfo{person}{Mihir
  Kshirsagar}, {and} \bibinfo{person}{Jonathan Mayer}.}
  \bibinfo{year}{2021}\natexlab{}.
\newblock \showarticletitle{What makes a dark pattern... dark? Design
  attributes, normative considerations, and measurement methods}. In
  \bibinfo{booktitle}{\emph{Proceedings of the 2021 CHI Conference on Human
  Factors in Computing Systems}}. \bibinfo{pages}{1--18}.
\newblock


\bibitem[\protect\citeauthoryear{Mathur, Narayanan, and Chetty}{Mathur
  et~al\mbox{.}}{2018}]%
        {mathur2018endorsements}
\bibfield{author}{\bibinfo{person}{Arunesh Mathur}, \bibinfo{person}{Arvind
  Narayanan}, {and} \bibinfo{person}{Marshini Chetty}.}
  \bibinfo{year}{2018}\natexlab{}.
\newblock \showarticletitle{Endorsements on social media: An empirical study of
  affiliate marketing disclosures on YouTube and Pinterest}.
\newblock \bibinfo{journal}{\emph{Proceedings of the ACM on Human-Computer
  Interaction}} \bibinfo{volume}{2}, \bibinfo{number}{CSCW}
  (\bibinfo{year}{2018}), \bibinfo{pages}{1--26}.
\newblock


\bibitem[\protect\citeauthoryear{Mittone and Savadori}{Mittone and
  Savadori}{2009}]%
        {mittone2009scarcity}
\bibfield{author}{\bibinfo{person}{Luigi Mittone} {and} \bibinfo{person}{Lucia
  Savadori}.} \bibinfo{year}{2009}\natexlab{}.
\newblock \showarticletitle{The scarcity bias}.
\newblock \bibinfo{journal}{\emph{Applied Psychology}} \bibinfo{volume}{58},
  \bibinfo{number}{3} (\bibinfo{year}{2009}), \bibinfo{pages}{453--468}.
\newblock


\bibitem[\protect\citeauthoryear{Moser, Schoenebeck, and Resnick}{Moser
  et~al\mbox{.}}{2019}]%
        {moser2019impulse}
\bibfield{author}{\bibinfo{person}{Carol Moser}, \bibinfo{person}{Sarita~Y
  Schoenebeck}, {and} \bibinfo{person}{Paul Resnick}.}
  \bibinfo{year}{2019}\natexlab{}.
\newblock \showarticletitle{Impulse Buying: Design Practices and Consumer
  Needs}. In \bibinfo{booktitle}{\emph{Proceedings of the 2019 CHI Conference
  on Human Factors in Computing Systems}}. \bibinfo{pages}{1--15}.
\newblock


\bibitem[\protect\citeauthoryear{Narayanan, Mathur, Chetty, and
  Kshirsagar}{Narayanan et~al\mbox{.}}{2020}]%
        {narayanan2020dark}
\bibfield{author}{\bibinfo{person}{Arvind Narayanan}, \bibinfo{person}{Arunesh
  Mathur}, \bibinfo{person}{Marshini Chetty}, {and} \bibinfo{person}{Mihir
  Kshirsagar}.} \bibinfo{year}{2020}\natexlab{}.
\newblock \showarticletitle{Dark Patterns: Past, Present, and Future: The
  evolution of tricky user interfaces}.
\newblock \bibinfo{journal}{\emph{Queue}} \bibinfo{volume}{18},
  \bibinfo{number}{2} (\bibinfo{year}{2020}), \bibinfo{pages}{67--92}.
\newblock


\bibitem[\protect\citeauthoryear{Ng}{Ng}{2013}]%
        {ng2013intention}
\bibfield{author}{\bibinfo{person}{Celeste See-Pui Ng}.}
  \bibinfo{year}{2013}\natexlab{}.
\newblock \showarticletitle{Intention to purchase on social commerce websites
  across cultures: A cross-regional study}.
\newblock \bibinfo{journal}{\emph{Information \& management}}
  \bibinfo{volume}{50}, \bibinfo{number}{8} (\bibinfo{year}{2013}),
  \bibinfo{pages}{609--620}.
\newblock


\bibitem[\protect\citeauthoryear{Ning and Yang}{Ning and Yang}{2021}]%
        {zheng2021legal}
\bibfield{author}{\bibinfo{person}{Zheng Ning} {and} \bibinfo{person}{Ge
  Yang}.} \bibinfo{year}{2021}\natexlab{}.
\newblock \showarticletitle{Legal Regulation and Improvement Path of E-commerce
  Live Broadcast}.
\newblock \bibinfo{journal}{\emph{Social Governance Review}}
  \bibinfo{number}{3} (\bibinfo{year}{2021}), \bibinfo{pages}{7}.
\newblock


\bibitem[\protect\citeauthoryear{Nodder}{Nodder}{2013}]%
        {nodder2013evil}
\bibfield{author}{\bibinfo{person}{Chris Nodder}.}
  \bibinfo{year}{2013}\natexlab{}.
\newblock \bibinfo{booktitle}{\emph{Evil by design: Interaction design to lead
  us into temptation}}.
\newblock \bibinfo{publisher}{John Wiley \& Sons}.
\newblock


\bibitem[\protect\citeauthoryear{{O'Ratings}}{{O'Ratings}}{2020}]%
        {oratings2020live}
\bibfield{author}{\bibinfo{person}{{O'Ratings}}.}
  \bibinfo{year}{2020}\natexlab{}.
\newblock \bibinfo{title}{O'Ratings Livestreaming User Observation Report}.
\newblock
\newblock
\urldef\tempurl%
\url{https://m.zol.com.cn/article/7505181.html}
\showURL{%
Retrieved 7/11/2021 from \tempurl}


\bibitem[\protect\citeauthoryear{Pellicone and Ahn}{Pellicone and Ahn}{2017}]%
        {pellicone2017game}
\bibfield{author}{\bibinfo{person}{Anthony~J. Pellicone} {and}
  \bibinfo{person}{June Ahn}.} \bibinfo{year}{2017}\natexlab{}.
\newblock \bibinfo{booktitle}{\emph{The Game of Performing Play: Understanding
  Streaming as Cultural Production}}.
\newblock \bibinfo{publisher}{Association for Computing Machinery},
  \bibinfo{address}{New York, NY, USA}, \bibinfo{pages}{4863–4874}.
\newblock
\showISBNx{9781450346559}
\urldef\tempurl%
\url{https://doi.org/10.1145/3025453.3025854}
\showURL{%
\tempurl}


\bibitem[\protect\citeauthoryear{Pires and Simon}{Pires and Simon}{2015}]%
        {pires2015youtube}
\bibfield{author}{\bibinfo{person}{Karine Pires} {and} \bibinfo{person}{Gwendal
  Simon}.} \bibinfo{year}{2015}\natexlab{}.
\newblock \showarticletitle{YouTube Live and Twitch: A Tour of User-Generated
  Live Streaming Systems}. In \bibinfo{booktitle}{\emph{Proceedings of the 6th
  ACM Multimedia Systems Conference}} (Portland, Oregon)
  \emph{(\bibinfo{series}{MMSys '15})}. \bibinfo{publisher}{Association for
  Computing Machinery}, \bibinfo{address}{New York, NY, USA},
  \bibinfo{pages}{225–230}.
\newblock
\showISBNx{9781450333511}
\urldef\tempurl%
\url{https://doi.org/10.1145/2713168.2713195}
\showDOI{\tempurl}


\bibitem[\protect\citeauthoryear{Scheibe, Fietkiewicz, and Stock}{Scheibe
  et~al\mbox{.}}{2016}]%
        {scheibe2016information}
\bibfield{author}{\bibinfo{person}{Katrin Scheibe}, \bibinfo{person}{Kaja~J
  Fietkiewicz}, {and} \bibinfo{person}{Wolfgang~G Stock}.}
  \bibinfo{year}{2016}\natexlab{}.
\newblock \showarticletitle{Information behavior on social live streaming
  services}.
\newblock  (\bibinfo{year}{2016}).
\newblock


\bibitem[\protect\citeauthoryear{Shanmugam, Sun, Amidi, Khani, and
  Khani}{Shanmugam et~al\mbox{.}}{2016}]%
        {shanmugam2016applications}
\bibfield{author}{\bibinfo{person}{Mohana Shanmugam}, \bibinfo{person}{Shiwei
  Sun}, \bibinfo{person}{Asra Amidi}, \bibinfo{person}{Farzad Khani}, {and}
  \bibinfo{person}{Fariborz Khani}.} \bibinfo{year}{2016}\natexlab{}.
\newblock \showarticletitle{The applications of social commerce constructs}.
\newblock \bibinfo{journal}{\emph{International Journal of Information
  Management}} \bibinfo{volume}{36}, \bibinfo{number}{3}
  (\bibinfo{year}{2016}), \bibinfo{pages}{425--432}.
\newblock


\bibitem[\protect\citeauthoryear{Sher and Su}{Sher and Su}{2019}]%
        {sher2019speedrunning}
\bibfield{author}{\bibinfo{person}{Stephen Tsung-Han Sher} {and}
  \bibinfo{person}{Norman~Makoto Su}.} \bibinfo{year}{2019}\natexlab{}.
\newblock \showarticletitle{Speedrunning for Charity: How Donations Gather
  Around a Live Streamed Couch}.
\newblock \bibinfo{journal}{\emph{Proc. ACM Hum.-Comput. Interact.}}
  \bibinfo{volume}{3}, \bibinfo{number}{CSCW}, Article \bibinfo{articleno}{48}
  (\bibinfo{date}{Nov.} \bibinfo{year}{2019}), \bibinfo{numpages}{26}~pages.
\newblock
\urldef\tempurl%
\url{https://doi.org/10.1145/3359150}
\showDOI{\tempurl}


\bibitem[\protect\citeauthoryear{Stephens, Hill, and Bergman}{Stephens
  et~al\mbox{.}}{1996}]%
        {stephens1996enhancing}
\bibfield{author}{\bibinfo{person}{Debra~Lynn Stephens},
  \bibinfo{person}{Ronald~Paul Hill}, {and} \bibinfo{person}{Karyn Bergman}.}
  \bibinfo{year}{1996}\natexlab{}.
\newblock \showarticletitle{Enhancing the consumer-product relationship:
  Lessons from the QVC home shopping channel}.
\newblock \bibinfo{journal}{\emph{Journal of business research}}
  \bibinfo{volume}{37}, \bibinfo{number}{3} (\bibinfo{year}{1996}),
  \bibinfo{pages}{193--200}.
\newblock


\bibitem[\protect\citeauthoryear{Strauss and Corbin}{Strauss and
  Corbin}{1994}]%
        {strauss1994grounded}
\bibfield{author}{\bibinfo{person}{Anselm Strauss} {and}
  \bibinfo{person}{Juliet Corbin}.} \bibinfo{year}{1994}\natexlab{}.
\newblock \showarticletitle{Grounded theory methodology: An overview.}
\newblock  (\bibinfo{year}{1994}).
\newblock


\bibitem[\protect\citeauthoryear{Su}{Su}{2019}]%
        {su2019empirical}
\bibfield{author}{\bibinfo{person}{Xiumei Su}.}
  \bibinfo{year}{2019}\natexlab{}.
\newblock \showarticletitle{An Empirical Study on the Influencing Factors of
  E-Commerce Live Streaming}. In \bibinfo{booktitle}{\emph{2019 International
  Conference on Economic Management and Model Engineering (ICEMME)}}. IEEE,
  \bibinfo{pages}{492--496}.
\newblock


\bibitem[\protect\citeauthoryear{Taylor}{Taylor}{2018}]%
        {taylor2018watch}
\bibfield{author}{\bibinfo{person}{TL Taylor}.}
  \bibinfo{year}{2018}\natexlab{}.
\newblock \bibinfo{booktitle}{\emph{Watch me play: Twitch and the rise of game
  live streaming}}.
\newblock \bibinfo{publisher}{Princeton University Press}.
\newblock


\bibitem[\protect\citeauthoryear{Waldman}{Waldman}{2020}]%
        {waldman2020cognitive}
\bibfield{author}{\bibinfo{person}{Ari~Ezra Waldman}.}
  \bibinfo{year}{2020}\natexlab{}.
\newblock \showarticletitle{Cognitive biases, dark patterns, and the `privacy
  paradox'}.
\newblock \bibinfo{journal}{\emph{Current opinion in psychology}}
  \bibinfo{volume}{31} (\bibinfo{year}{2020}), \bibinfo{pages}{105--109}.
\newblock


\bibitem[\protect\citeauthoryear{Warner and Fischer}{Warner and
  Fischer}{2019}]%
        {warner2019senators}
\bibfield{author}{\bibinfo{person}{Mark Warner} {and} \bibinfo{person}{Debra
  Fischer}.} \bibinfo{year}{2019}\natexlab{}.
\newblock \bibinfo{title}{Senators Introduce Bipartisan Legislation to Ban
  Manipulative `Dark Patterns'}.
\newblock
\newblock
\urldef\tempurl%
\url{https://www.fischer.senate.gov/public/index.cfm/2019/4/senators-introduce-bipartisan-legislation-to-ban-manipulative-dark-patterns}
\showURL{%
Retrieved 6/19/2022 from \tempurl}


\bibitem[\protect\citeauthoryear{Wohn}{Wohn}{2019}]%
        {wohn2019volunteer}
\bibfield{author}{\bibinfo{person}{Donghee~Yvette Wohn}.}
  \bibinfo{year}{2019}\natexlab{}.
\newblock \showarticletitle{Volunteer Moderators in Twitch Micro Communities:
  How They Get Involved, the Roles They Play, and the Emotional Labor They
  Experience}. In \bibinfo{booktitle}{\emph{Proceedings of the 2019 CHI
  Conference on Human Factors in Computing Systems}} (Glasgow, Scotland Uk)
  \emph{(\bibinfo{series}{CHI '19})}. \bibinfo{publisher}{Association for
  Computing Machinery}, \bibinfo{address}{New York, NY, USA},
  \bibinfo{pages}{1–13}.
\newblock
\showISBNx{9781450359702}
\urldef\tempurl%
\url{https://doi.org/10.1145/3290605.3300390}
\showDOI{\tempurl}


\bibitem[\protect\citeauthoryear{Wohn, Freeman, and McLaughlin}{Wohn
  et~al\mbox{.}}{2018}]%
        {wohn2018explaining}
\bibfield{author}{\bibinfo{person}{Donghee~Yvette Wohn}, \bibinfo{person}{Guo
  Freeman}, {and} \bibinfo{person}{Caitlin McLaughlin}.}
  \bibinfo{year}{2018}\natexlab{}.
\newblock \showarticletitle{Explaining Viewers' Emotional, Instrumental, and
  Financial Support Provision for Live Streamers}. In
  \bibinfo{booktitle}{\emph{Proceedings of the 2018 CHI Conference on Human
  Factors in Computing Systems}} (Montreal QC, Canada)
  \emph{(\bibinfo{series}{CHI '18})}. \bibinfo{publisher}{Association for
  Computing Machinery}, \bibinfo{address}{New York, NY, USA},
  \bibinfo{pages}{1–13}.
\newblock
\showISBNx{9781450356206}
\urldef\tempurl%
\url{https://doi.org/10.1145/3173574.3174048}
\showDOI{\tempurl}


\bibitem[\protect\citeauthoryear{Wongkitrungrueng, Dehouche, and
  Assarut}{Wongkitrungrueng et~al\mbox{.}}{2020}]%
        {wongkitrungrueng2020live}
\bibfield{author}{\bibinfo{person}{Apiradee Wongkitrungrueng},
  \bibinfo{person}{Nassim Dehouche}, {and} \bibinfo{person}{Nuttapol Assarut}.}
  \bibinfo{year}{2020}\natexlab{}.
\newblock \showarticletitle{Live streaming commerce from the sellers'
  perspective: implications for online relationship marketing}.
\newblock \bibinfo{journal}{\emph{Journal of Marketing Management}}
  \bibinfo{volume}{36}, \bibinfo{number}{5-6} (\bibinfo{year}{2020}),
  \bibinfo{pages}{488--518}.
\newblock


\bibitem[\protect\citeauthoryear{Wu}{Wu}{2016}]%
        {wu2016youtube}
\bibfield{author}{\bibinfo{person}{Katrina Wu}.}
  \bibinfo{year}{2016}\natexlab{}.
\newblock \showarticletitle{YouTube marketing: Legality of sponsorship and
  endorsements in advertising}.
\newblock \bibinfo{journal}{\emph{JL Bus. \& Ethics}}  \bibinfo{volume}{22}
  (\bibinfo{year}{2016}), \bibinfo{pages}{59}.
\newblock


\bibitem[\protect\citeauthoryear{Wu, Sang, Zhang, and Huang}{Wu
  et~al\mbox{.}}{2018}]%
        {wu2018danmaku}
\bibfield{author}{\bibinfo{person}{Qunfang Wu}, \bibinfo{person}{Yisi Sang},
  \bibinfo{person}{Shan Zhang}, {and} \bibinfo{person}{Yun Huang}.}
  \bibinfo{year}{2018}\natexlab{}.
\newblock \showarticletitle{Danmaku vs. forum comments: understanding user
  participation and knowledge sharing in online videos}. In
  \bibinfo{booktitle}{\emph{Proceedings of the 2018 ACM conference on
  supporting groupwork}}. \bibinfo{pages}{209--218}.
\newblock


\bibitem[\protect\citeauthoryear{Xu, Wu, Chang, and Li}{Xu
  et~al\mbox{.}}{2019}]%
        {xu2019investigation}
\bibfield{author}{\bibinfo{person}{Xiaoyu Xu}, \bibinfo{person}{Jen-Her Wu},
  \bibinfo{person}{Ya-Ting Chang}, {and} \bibinfo{person}{Qi Li}.}
  \bibinfo{year}{2019}\natexlab{}.
\newblock \showarticletitle{The Investigation of Hedonic Consumption, Impulsive
  Consumption and Social Sharing in E-commerce Live-streaming Videos.}. In
  \bibinfo{booktitle}{\emph{PACIS}}. \bibinfo{pages}{43}.
\newblock


\bibitem[\protect\citeauthoryear{Xu, Wu, and Li}{Xu et~al\mbox{.}}{2020}]%
        {xu2020drives}
\bibfield{author}{\bibinfo{person}{Xiaoyu Xu}, \bibinfo{person}{Jen-Her Wu},
  {and} \bibinfo{person}{Qi Li}.} \bibinfo{year}{2020}\natexlab{}.
\newblock \showarticletitle{What Drives Consumer Shopping Behavior in Live
  Streaming Commerce?}
\newblock \bibinfo{journal}{\emph{Journal of Electronic Commerce Research}}
  \bibinfo{volume}{21}, \bibinfo{number}{3} (\bibinfo{year}{2020}),
  \bibinfo{pages}{144--167}.
\newblock


\bibitem[\protect\citeauthoryear{Yu and Goh}{Yu and Goh}{2020}]%
        {reuters2020livestreaming}
\bibfield{author}{\bibinfo{person}{Sophie Yu} {and} \bibinfo{person}{Brenda
  Goh}.} \bibinfo{year}{2020}\natexlab{}.
\newblock \bibinfo{title}{China drafts rules to govern its booming
  livestreaming sales industry}.
\newblock
\newblock
\urldef\tempurl%
\url{https://www.reuters.com/article/us-china-internet-regulation-idUSKBN27T1B0}
\showURL{%
Retrieved 7/6/2021 from \tempurl}


\bibitem[\protect\citeauthoryear{Zagal, Bj{\"o}rk, and Lewis}{Zagal
  et~al\mbox{.}}{2013}]%
        {zagal2013dark}
\bibfield{author}{\bibinfo{person}{Jos{\'e}~P Zagal}, \bibinfo{person}{Staffan
  Bj{\"o}rk}, {and} \bibinfo{person}{Chris Lewis}.}
  \bibinfo{year}{2013}\natexlab{}.
\newblock \showarticletitle{Dark patterns in the design of games}.
\newblock  (\bibinfo{year}{2013}).
\newblock


\bibitem[\protect\citeauthoryear{Zhu, Yang, and Dai}{Zhu et~al\mbox{.}}{2017}]%
        {zhu2017understanding}
\bibfield{author}{\bibinfo{person}{Zhenhui Zhu}, \bibinfo{person}{Zhi Yang},
  {and} \bibinfo{person}{Yafei Dai}.} \bibinfo{year}{2017}\natexlab{}.
\newblock \showarticletitle{Understanding the gift-sending interaction on
  live-streaming video websites}. In \bibinfo{booktitle}{\emph{International
  Conference on social computing and social media}}. Springer,
  \bibinfo{pages}{274--285}.
\newblock


\end{thebibliography}


\end{CJK}
\end{document}